\documentclass[twocolumn]{aastex631}
\usepackage{CJK}
\usepackage{verbatim}
\usepackage{amsmath}
\usepackage{soul, xcolor}
\usepackage{booktabs}

\newcommand{\RNum}[1]{\uppercase\expandafter{\romannumeral #1\relax}}

\shorttitle{Off-centered Extended Emission of the Little Red Dots}
\shortauthors{Chen et al.}

\begin{document}
\begin{CJK*}{UTF8}{gbsn}

\title{The Physical Nature of the Off-centered Extended Emission Associated with the Little Red Dots}

\author[0009-0003-4721-177X]{Chang-Hao Chen (陈昌灏)}
\affiliation{Kavli Institute for Astronomy and Astrophysics, Peking University, Beijing 100871, China}
\affiliation{Department of Astronomy, School of Physics, Peking University, Beijing 100871, China}

\author[0000-0001-6947-5846]{Luis C. Ho}
\affiliation{Kavli Institute for Astronomy and Astrophysics, Peking University, Beijing 100871, China}
\affiliation{Department of Astronomy, School of Physics, Peking University, Beijing 100871, China}

\author[0000-0001-8496-4162]{Ruancun Li (李阮存)}
\affiliation{Kavli Institute for Astronomy and Astrophysics, Peking University, Beijing 100871, China}
\affiliation{Department of Astronomy, School of Physics, Peking University, Beijing 100871, China}

\author[0000-0001-9840-4959]{Kohei Inayoshi}
\affiliation{Kavli Institute for Astronomy and Astrophysics, Peking University, Beijing 100871, China}
\affiliation{Department of Astronomy, School of Physics, Peking University, Beijing 100871, China}

\begin{abstract}

A significant fraction of little red dots (LRDs) exhibit nearby extended emission of unknown origin. If physically associated with the LRD, this component may trace stellar emission from an off-centered host galaxy, neighboring companions, or nebular gas illuminated by the active nucleus. We investigate the detailed spectral energy distribution of the extended emission near four LRDs in the JWST UNCOVER and MegaScience surveys. We accurately decompose the extended emission from the dominant point source by simultaneously fitting the images in eight broad-band and nine medium-band filters. After considering both the results from photometric redshift fitting and the probability of galaxies at different redshift overlapping, we confirm that the off-centered blobs in three sources are physically associated with the LRDs, with two of them showing strong [\ion{O}{3}] $\lambda\lambda 4959,\,5007$ emission captured by the medium-band filters. While the spectral energy distributions of all three blobs can be modeled assuming star-forming galaxies with stellar mass $\sim 10^8\,M_{\odot}$, the exceptionally strong [\ion{O}{3}] emission of two sources is best interpreted as pure nebular emission from low-density ($n<10\, {\rm cm}^{-3}$), low-metallicity ($Z\approx 0.05\,Z_{\odot}$) gas photoionized by the ultraviolet radiation from the nearby LRD. Adopting LRD halo masses constrained by clustering measurements and theoretical considerations, we estimate a typical baryonic halo mass accretion rate of $\sim 2-9\, M_{\odot}\,{\rm yr}^{-1}$. If the halo accretion rate is sustained to $z=4$ and stars form with an efficiency of 10\%, the accreted gas would form a galaxy with stellar mass $\sim 10^9\,M_{\odot}$, potentially rendering them spatially resolved at lower redshift.
\end{abstract}


\section{Introduction} \label{intro}

While searching the early Universe for supermassive black holes in their infancy, the James Webb Space Telescope (JWST) came across numerous point-like sources with distinctive colors, a population now popularly dubbed the little red dots (LRDs). These objects are often observed to have a characteristic ``V-shape'' spectral energy distribution (SED) that is blue in the rest-frame ultraviolet (UV) and red in the optical, as well as extremely compact physical size of $\sim 100\, {\rm pc}$ or less \citep{Kocevski2023, Furtak2024, Kokorev2024, Matthee2024, Akins2025, Kocevski2025, Labbe2025}. The red continuum slope of the LRDs flattens toward the near- and mid-infrared \citep{Perez-Gonzalez2024, Williams2024, Akins2025, Naidu2025, Wang2025} and eventually becomes completely undetected in the far-infrared \citep{Akins2025, Labbe2025, Setton2025}. Apart from establishing their extragalactic nature and high redshifts ($z \approx 4-8$), follow-up spectroscopic observations reveal that LRDs possesses strong ${\rm H}\alpha$ and ${\rm H}\beta$ emission lines with full width at half maximum $\gtrsim 1000\, {\rm km\,s^{-1}}$ \citep{Greene2024}, superficialy similar to those seen in broad-line active galactic nuclei (AGNs). If local estimators of virial mass (e.g., \citealt{Greene2005}) can be applied, LRDs have estimated black holes masses $\sim 10^6-10^8\,M_\odot$ radiating at Eddington ratios of $\sim 0.1-1$. 

The physical nature of the LRDs---whether they are early-formed galaxies or AGNs---is a subject of intensive ongoing debate. Under the galaxy scenario, a highly dust-reddened stellar population produces the red optical slope as well as the prominent drop near $\sim 4000$ \AA\ that resembles the Balmer break observed in relatively evolved stellar populations \citep{Furtak2024, Greene2024, Wang2024, Wang2025}. On the other hand, while it is possible to generate broad emission lines in extremely dense stellar systems \citep{Baggen2024}, such systems appear to be dynamically unstable to runaway collisions \citep{Guia2024}. Forming such massive stellar systems at early cosmic epoch would also be in conflict with the standard framework of structure formation \citep{Inayoshi2024}. Under the AGN scenario, emission from the accretion disk needs to be attenuated by a combination of dust and high-density ($10^9-10^{11}\,{\rm cm}^{-3}$) gas in order to simultaneously produce both the red optical slope and Balmer absorption features \citep{Inayoshi2025b, Ji2025, Naidu2025}. Although the degree of dust attenuation varies with different models, dust geometry, and possibly grain size distribution \citep{Killi2024, Li2025}, all current models struggle to explain the deficit of dust emission \citep{Labbe2025, Setton2025}.

Besides the already enchanting story for the ``dots'' themselves, another interesting fact often left undiscussed in the literature is that many LRDs exhibit nearby emission of unknown origin. \citet{Chen2025} discovered extended, off-centered emission near the dominant point source in five out of eight LRDs. A more robust statistic is later reported in \citet{Rinaldi2025}, in which 30 out of 99 LRDs display complex UV morphologies, featuring either a highly symmetric structure or at least two distinctive neighboring sources. In a few cases in which detailed parametric morphologies are measured for these extended blobs, they are found to have low central concentration, characterized by a S\'ersic profile with $n\leq 1$ \citep{Chen2025, Labbe2025b}.

The number density of LRDs is significantly higher than UV luminous quasars at similar redshift \citep{Greene2024, Kokorev2024, Matthee2024}, which implies that most LRDs will not become quasars but end up as normal Seyferts or composite (AGN plus star-forming) galaxies in the local Universe. As the only resolvable structures near the LRDs, these off-centered blobs are an important piece of the puzzle as to how the LRDs gradually lose their characteristic signatures and join the general galaxy population at lower redshifts. Understand the physical properties of these blobs, as well as their potential interactions with the LRDs, would shed light into the evolution of these peculiar objects. The small angular separation between these blobs and the LRDs render their emission heavily blended and in some cases almost completely dominated by the light from the point source. As a result, deriving the SED for the off-centered emission is particularly challenging. \citet{Chen2025} made a first attempt with images in six broad-band filters and one medium-band filter. By simultaneously decomposing the images using parametric models, they found that the SEDs of three off-centered blobs in their sample can be modeled by stellar populations at similar or lower redshift as the LRD point sources, while the SEDs of the blobs in the other LRDs cannot be well-fit with either galaxy or nebular gas ionized by stars. A more extreme case has been reported in \cite{Tanaka2025}, in which the nearby source has similar SED as the predominant point source, prompting the authors to regard them as dual LRDs.

This Letter revisits four LRDs previously analyzed in \citet{Chen2025}. By simultaneously analyzing images in eight broad-band filters supplemented with nine additional medium-band filters, we offer a much more precise depiction of the SEDs of the off-centered blobs, hoping to better unveil their nature. We assume a flat cosmology with $\Omega_{m}=0.3111$, $\Omega_{\Lambda}=0.6889$, and Hubble constant $H_0=67.66\,{\rm km}\,{\rm s}^{-1}\, {\rm Mpc}^{-1}$ \citep{Planck2018}.

\section{Data and Sample} \label{data_sample}

The data used in this work were taken as part of the JWST Cycle~1 Treasury program Ultradeep NIRSpec and NIRCam ObserVations before the Epoch of Reionization (UNCOVER; \citealt{Bezanson2024}), and the Cycle~2 Medium Bands, Mega Science (MegaScience; \citealt{Suess2024}) survey. Together, the UNCOVER and MegaScience surveys provide a comprehensive set of deep ($\sim 28-30\, {\rm AB~mag}$) images in \textit{all} NIRCam medium-band and broad-band filters spanning from F070W to F480M, centered on the galaxy cluster Abell~2744 at $z=0.308$. We utilize the data release 3 (DR3) multi-band mosaic images from the UNCOVER and MegaScience reduced by \citet{Bezanson2024} and \citet{Suess2024}, respectively, in which the images were aligned to stars from the Gaia DR3 catalog \citep{Gaia2023}, co-added, and subtracted for the large-scale background using the {\tt\string Grizli} pipeline \citep{Brammer2022}. Images in the short-wavelength (SW) bands (F070W, F090W, F115W, F140M, F150W, F162M, F182M, F200W, and F210M) were drizzled to a pixel scale of 0\farcs02, while those in the long-wavelength (LW) bands (F250M, F277W, F300M, F335M, F356W, F360M, F410M, F430M, F444W, F460M, and F480M) were drizzled to a pixel scale of 0\farcs04. The final data products include mosaics from 20 filters.

The targets analyzed in this work (MSA10686, MSA13821, MSA41225, MSA45924) were initially classified as LRDs in \citet{Labbe2025} based on the identification of their point source-dominated morphology exhibiting a characteristic ``V-shape'' SED. Follow-up observations with NIRSpec/PRISM using the Micro-Shutter Assembly (MSA) were subsequently conducted by \citet{Greene2024}, who reported (tentative) detection of broad H$\alpha$ emission. The simultaneous seven-band (F115W$-$F444W) image analysis by \citet{Chen2025} revealed extended off-centered components located within 0\farcs2 ($\sim 1\,\rm kpc$) of the dominant point sources\footnote{\citet{Chen2025} also discovered an off-centered blob in another source, MSA4286, but we removed it from our sample since it has not been imaged by the MegaScience program.}. The SED analysis of Chen et al. suggests that the off-centered blob near MSA41225 could be produced by a projected, intervening galaxy at $z\approx 2.7$, while the off-centered emission associated with MSA10686, MSA13821, and MSA45924 is consistent with either either a nearby galaxy or nebular emission at the same redshift as the point sources.

\section{Image Fitting} \label{image_fitting}

\subsection{PSF Construction} \label{PSF}

We follow the procedures in \citet{Chen2025} for PSF model construction. For each target, field stars observed with the same telescope on-sky position angle are selected by identifying sources with point-like morphology using {\tt\string SExtractor} \citep{Bertin1996} and comparing their SEDs with stellar spectral templates. Empirical PSF models are then generated using the pixel-based algorithm {\tt\string PSFEx} \citep{Bertin2013}. The signal-to-noise ratio of the field-star images varies with wavelength, as the LW filters generally have lower signal-to-noise ratio compared to the SW filters because of the decline in stellar flux toward the redder bands. The signal-to-noise ratio of the medium-band filters is also lower that that of the broad-band filters owing to their narrower bandpass. In order to achieve PSFs with more uniform quality across different bands, we construct hybrid PSF models that combine the central core of an empirical PSF with the outer wings of a theoretical PSF generated using {\tt\string WebbPSF} \citep{Perrin2015}. For the two reddest medium bands (F460M and F480M), even the inner core of their PSFs proved impossible to constrain, and we therefore excluded them from further consideration.

\begin{figure*}[ht!]
\centering
\includegraphics[width=0.49\textwidth]{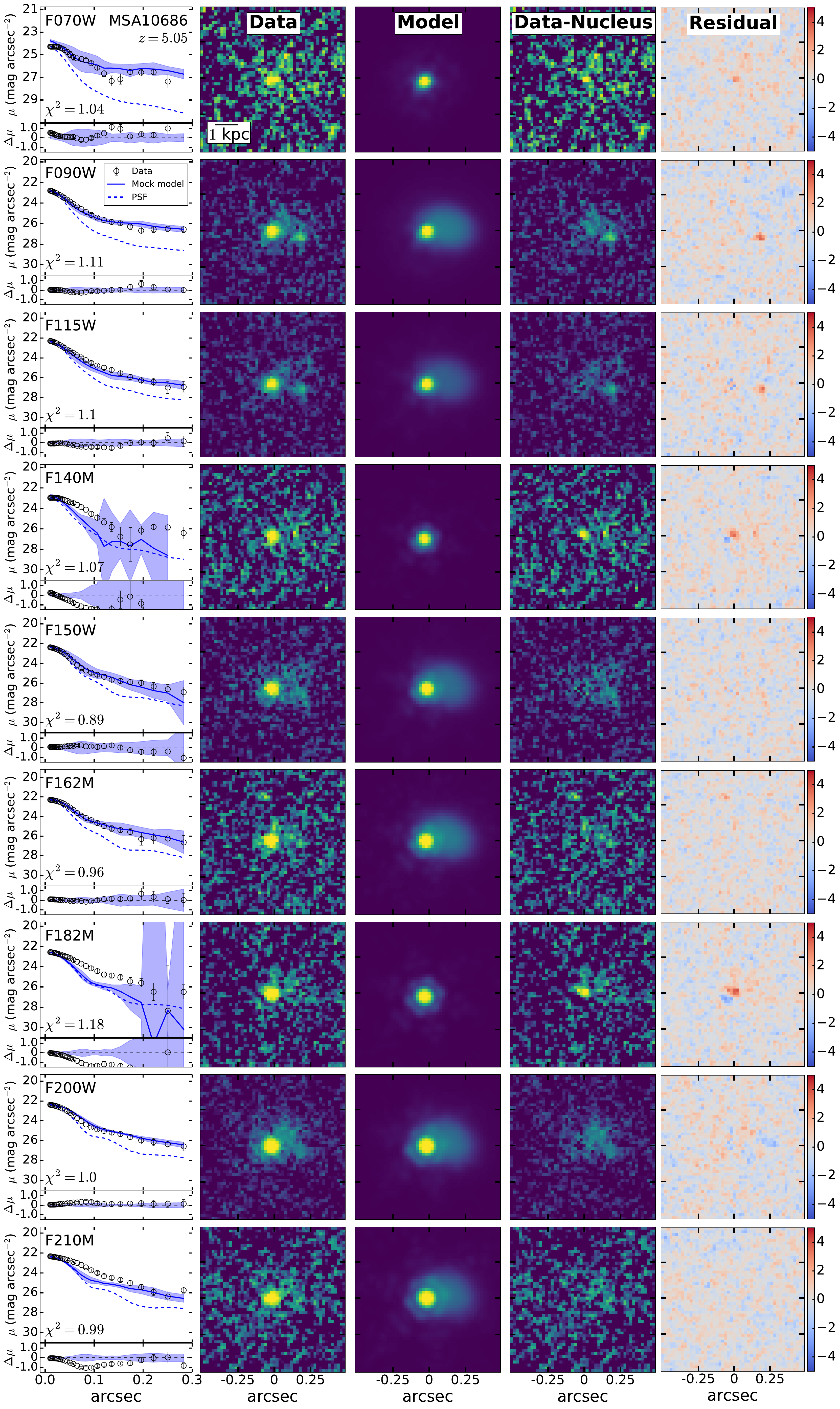}
\includegraphics[width=0.49\textwidth]{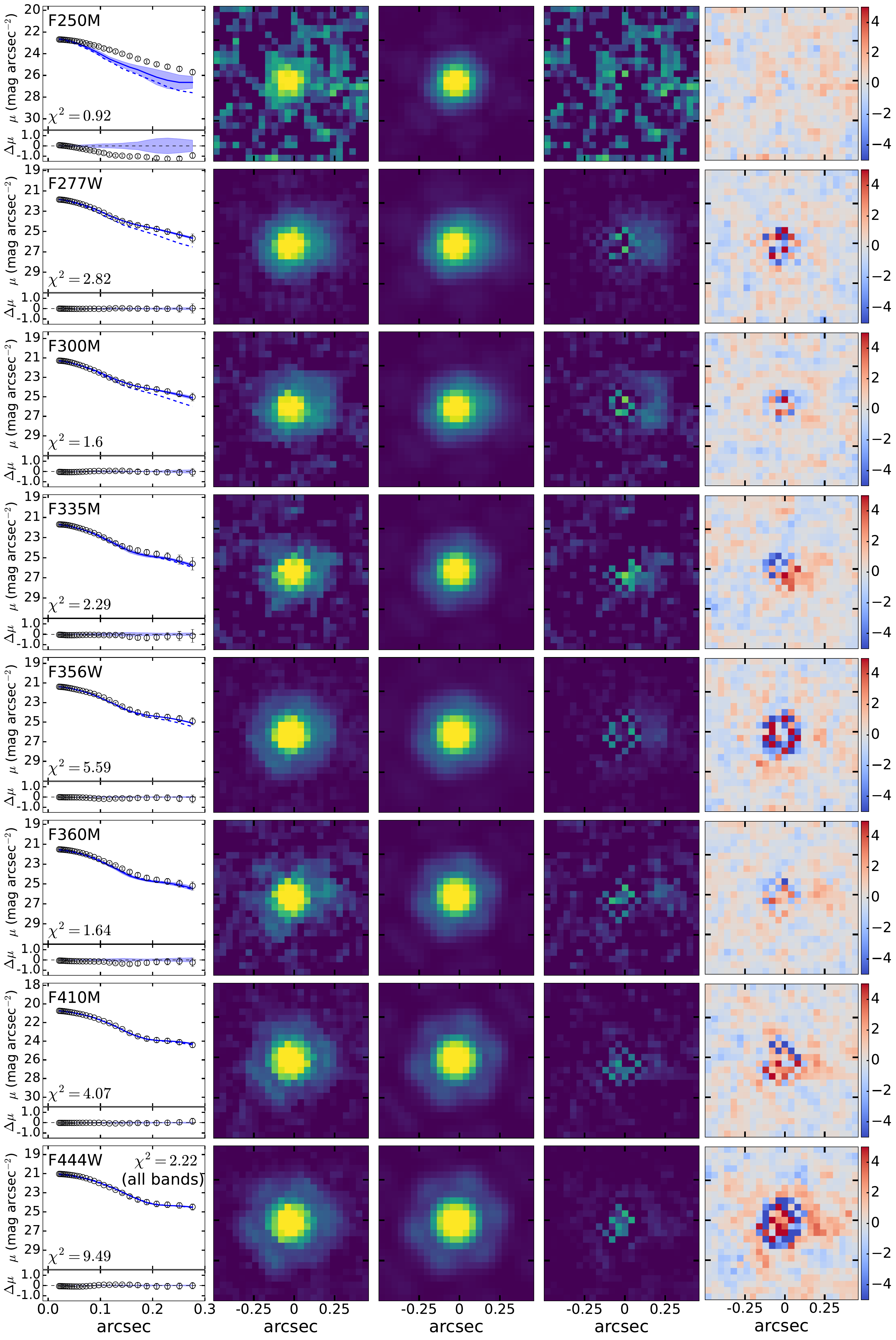}
\caption{Simultaneous multi-band image fitting results for MSA10686 at $z=5.05$. Each row shows the result for one NIRCam band. In the left-most column, the upper panel of each row shows the radial surface brightness distribution (open circles with error bars), the PSF model (blue dotted line), as well as the median profile of the best-fit model (blue solid line) and its standard deviation (blue shaded region), derived by randomly inserting the model into empty sky regions within the field of view. We present the best-fit point source $+$ S\'ersic model for bands where the off-centered blobs are detected, and single point source model for bands where the blobs are not detected. The $\chi^2$ for each band is given in the lower-left corner of each panel. The lower subpanels give the residuals between the data and the best-fit model (data$-$model). The imaging columns, from left to right, display the original data, best-fit model, data minus the nucleus component, and residuals normalized by the errors (data$-$model$/$error), which are stretched linearly from $-5$ to 5.
\label{10686_fitting_results}}
\end{figure*}

\begin{deluxetable*}{cccccccccc}[]
\label{sw_sed}
\tabletypesize{\scriptsize}
\tablecaption{Short-wavelength Photometry of the Off-centered Emission}
\tablewidth{0pt}
\tablehead{MSA & F070W & F090W & F115W & F140M & F150W & F162M & F182M & F200W & F210M \\
                 & (nJy) & (nJy) & (nJy) & (nJy) & (nJy) & (nJy) & (nJy) & (nJy) & (nJy) }
\startdata
10686 & $<12.02$ & $21.26\pm1.41$ & $20.61\pm1.18$ & $<35.98$ & $24.45\pm1.48$ & $31.85\pm1.72$ & $<22.70$ & $26.98\pm1.60$ & $26.36\pm1.95$ \\
13821 & $<5.81$ & $<11.07$ & $9.31\pm1.59$ & $<18.71$ & $10.41\pm3.22$ & $<18.54$ & $<12.25$ & $9.18\pm1.25$ & $<12.47$ \\
41225 & $<7.18$ & $5.93\pm0.90$ & $8.78\pm1.78$ & $<17.06$ & $9.11\pm1.16$ & $15.62\pm2.32$ & $<11.38$ & $16.24\pm1.83$ & $12.35\pm2.11$ \\
45924 & $46.49\pm1.07$ & $57.88\pm1.11$ & $50.85\pm1.07$ & $53.41\pm1.02$ & $38.39\pm0.89$ & $34.17\pm0.81$ & $36.52\pm0.87$ & $47.84\pm1.23$ & $38.65\pm0.79$ \\
\enddata
\end{deluxetable*}

\begin{deluxetable*}{cccccccccc}[]
\label{lw_sed}
\tabletypesize{\scriptsize}
\tablecaption{Long-wavelength Photometry of the Off-centered Emission}
\tablewidth{0pt}
\tablehead{MSA & F250M & F277W & F300M & F335M & F356W & F360M & F410M & F444W \\
                 & (nJy) & (nJy) & (nJy) & (nJy) & (nJy) & (nJy) & (nJy) & (nJy) }
\startdata
10686 & $<42.86$ & $39.77\pm2.44$ & $56.10\pm3.33$ & $<21.28$ & $29.29\pm1.80$ & $<33.73$ & $<45.29$ & $<42.86$ \\
13821 & $<15.85$ & $<11.38$ & $17.50\pm3.54$ & $<8.95$ & $18.63\pm2.77$ & $31.45\pm4.23$ & $<7.94$ & $<13.43$ \\
41225 & $<21.28$ & $14.48\pm1.72$ & $19.72\pm2.29$ & $15.03\pm2.25$ & $13.76\pm1.80$ & $20.17\pm2.96$ & $<19.06$ & $12.16\pm2.96$ \\
45924 & $<124.75$ & $<78.71$ & $<122.47$ & $<135.53$ & $<197.71$ & $<313.35$ & $<197.71$ & $<197.71$ \\
\enddata
\end{deluxetable*}

\subsection{Data Preparation and Fitting Procedure}

We use {\tt\string GalfitS} (R. Li \& L. C. Ho, in preparation), a parametric image modeling code that simultaneously fits multi-band images with different resolutions and pixel scales. For each source, the fit is conducted on a $1\arcsec \times 1\arcsec$ image cutout centered on the point source. The input error map contains three major components: the background, readout noise provided by the weight maps in UNCOVER DR3, and Poisson noise estimated from the mosaics. The final error map is generated adopting the procedures described in the DAWN JWST Archive\footnote{\url{https://dawn-cph.github.io/dja/index.html}}. To remove nearby contaminants for MSA13821, we follow \citet{Chen2025} to create a $4\arcsec \times 4\arcsec$ cutout image in the F277W band that serves as the detection image, perform source detection using the {\tt\string Photutils.detect\_source} package with {\tt\string nstd\,=\,0.7, npix\,=\,15}, and {\tt\string contrast\,=\,0.01}, and then measure and remove the local sky background with all sources masked. For MSA10686 and MSA41225, extended wings of foreground galaxies are visible within their $1\arcsec \times 1\arcsec$ cutout images. As these potentially can affect the measurement for the extended off-centered component of the LRD, especially in the LW filters, we do not mask but simultaneously fit and subtract them from the cutout images in all bands. Specifically, we model the light profiles of the nearby galaxies with a \citet{Sersic1968} function and the LRDs are described as a point source. Then, the nearby galaxies, LRDs, and a uniform background model are fit together in cutout images large enough to include the centers of all nearby galaxies, which in our case is $2\arcsec \times 2\arcsec$ for MSA10686 and $4\arcsec \times 4\arcsec$ for MSA41225. During the fitting, $R_{e}$, $n$, and the positions of the S\'ersic and point source profiles are fixed across different bands, allowing only the brightness of the light profiles and background to vary. The extended wings from nearby galaxies as well as the local background in the final $1\arcsec \times 1\arcsec$ cutout images are subsequently subtracted using the best-fit models. Note that in the above fitting, we do not include an additional component to describe the off-centered blobs, because the morphology and brightness of these low-surface brightness components hardly can be constrained in the relatively large image cutouts. Since our goal in this initial step is to remove contamination from the extended wings of much brighter nearby galaxies, the effect of these much fainter blobs should be minimal. 

For the final fitting to characterize the off-centered blobs, we adopt the cutout images with background subtracted and nearby sources masked. The cutout size is $1\arcsec \times 1\arcsec$ for all targets except for the much brighter MSA45924, for which we use a $2\arcsec \times 2\arcsec$ cutout to include the prominent PSF diffraction spikes. We model the light profile of the LRD with a point source and adopt the S\'ersic profile to represent the off-centered blobs. Initial values of structural parameters are set according to the values provided in \citet{Chen2025}; the fit assumes that the parameters are constant in different bands. All the fits use the Adam Optimizer as the optimization algorithm to locate the position of maximum likelihood in parameter space. We run this process 20,000 times. We then perform nested sampling around the position of maximum likelihood to infer parameter uncertainties from the posterior distribution. Figure~\ref{10686_fitting_results} illustrates the fit for MSA10686. Results for the other three sources are shown in Appendix~A.

\subsection{Estimation of Flux Upper Limit}

To minimize potential bias in the SED modeling of the off-centered blobs, we adopt stricter constraints compare to \citet{Chen2025} in determining their flux upper limits. We first run a set of additional fits using only a point source to model the LRD itself. We then evaluate the goodness-of-fit in each band using the Bayesian information criterion,

\begin{equation}
{\rm BIC} = k\, {\rm ln}\,N -2\,{\rm ln}\,L,
\end{equation}

\noindent
with $k$ the number of model parameters, $N$ the number of data points, and $L$ the fitting likelihood. A smaller BIC value indicates better fitting performance. For a given band, if including the off-centered S\'ersic model does not reduce the BIC value by more than 10, we treat the off-centered blob as a non-detection and perform realistic mock simulations to place an upper limit on its flux. We begin by visually selecting empty sky regions from the UNCOVER mosaics that have the same on-sky position angle combination as the target, requiring that the surface brightness fluctuation level of the empty sky region differs by less than 0.5 mag ${\rm arcsec}^{-2}$ compared with the background near the target to ensure that the mock image is truly an analog of the observed data. We then generate a mock model that includes the best-fit point-source component as well as the best-fit off-centered S\'ersic component whose total flux has been rescaled between 26 to 30 mag, in steps of 0.5 mag. For each magnitude step, 10 mock images are generated, each by combining the sky background, mock model, and random pixel-to-pixel Poisson noise calculated from all the components. We then fit all the mock images with an off-centered S\'ersic $+$ point-source model, as well as a single point-source model, and compare the BIC values to calculate the detection fraction of the off-centered component in each magnitude bin. The $1\, \sigma$ flux upper limit is defined as the value where the detection probability reaches 68.2\%, with the exact number calculated by linearly interpolating between the two adjacent magnitude bins whose detection probabilities are below and above $1\, \sigma$. 

For the bands that satisfy the aforementioned BIC selection criterion, we further compare the measured flux of the off-centered blobs with the detection limit of the same band. The best-fit fluxes are only taken as solid detection if they surpass the detection limit by more than the uncertainty of the best-fit flux. We report the final photometric measurements in Tables~\ref{sw_sed} and \ref{lw_sed}.

\section{SED Analysis}\label{sed_analysis}

\subsection{Photometric Redshift and Overlapping Probability}\label{phot-z}

We first estimate photometric redshifts using {\tt\string EAzY} \citep{Brammer2008}, aiming to determine whether the off-centered blobs are physically associated with the LRD point sources, or just galaxies at different redshift projected close to the line-of-sight. We use the standard set of 12 templates in {\tt\string EAzY} generated from the stellar population synthesis code of \cite{Conroy2010}. To account for the possible high-redshift nature of these blobs, we also include templates from \citet{Larson2023} that are optimized for typical galaxies in the range $4\lesssim z \lesssim 7$. Upper limits during the fits are treated as data points with both their values and errors equal to 0.5 times the actual upper limits. This configuration allows the model values in these bands to vary between zero and the upper limit with the $1\, \sigma$ confidence, with preference given to model fluxes closer to one half of the detection upper limits. The consequences of this treatment of the upper limits will be explored further in Section~\ref{galaxy_model}. No magnitude priors are adopted. We found that all four sources exhibit bimodal redshift probability distribution functions, with one peak located near the spectroscopic redshift of the LRDs and the other at a lower redshift.

\begin{figure}[t]
\centering
\includegraphics[width=0.49\textwidth]{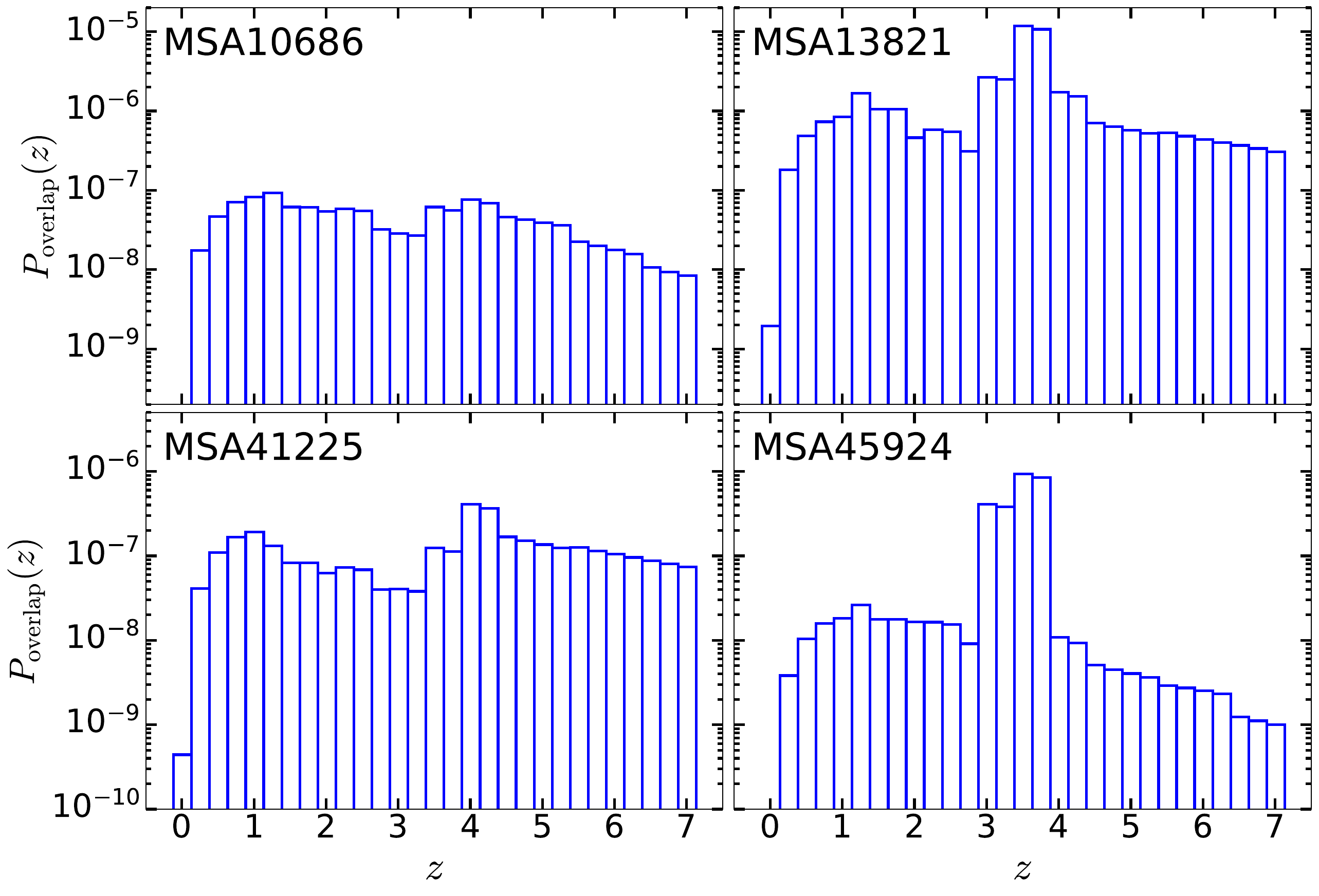}
\caption{Overlapping probability as a function of redshift for the off-centered emission.
\label{p_overlap}}
\end{figure}

To quantify the probability that the off-centered blobs lie at redshifts discordant with those of the LRDs, we follow the procedures in \citet{Bloom2002}, \citet{Berger2010}, and \citet{Wu2022}, assuming a uniform spatial distribution of galaxies such that the overlapping probability can be described by a Poisson distribution,

\begin{equation}
p(z)\, = \, 1\, - \, e^{-2\pi r_{d} \cdot \delta r_{d} \cdot \Delta D \cdot \Phi (z,m_0,\delta m_0)},
\end{equation}

\noindent
in which $r_{d}$ and $\delta r_{d}$ are the best-fit value and uncertainty of the distance between the point source and the center of the blob, $\Delta D$ is the line-of-sight comoving distance within a certain redshift bin, and $\Phi (z,m_0,\delta m_0)$ is the comoving number density of galaxies at redshift $z$, given the apparent magnitude $m_0$ and magnitude uncertainty $\delta m_0$ of the blob. The comoving number density is calculated as

\begin{equation}
\Phi (z,m_0,\delta m_0)=\int_{m_0-\delta m_0}^{m_0+\delta m_0} \phi(z,m)\, dm , 
\end{equation}

\noindent
where $\phi(z,m)$ is the galaxy luminosity function at redshift $z$. Given our wavelength coverage from 0.7 to 4.4 $\mu {m}$, we adopt the rest-frame $V$-band luminosity function for $0<z\leq 4$ and the rest-frame 1600\,\AA\ UV luminosity function for $4<z\leq 7$. Specifically, at $z\leq 0.4$ we use the $r$-band luminosity function from \citet{Loveday2012}, and at $0.4<z\leq 4$ we adopt the results from \citet{Marchesini2012}, who derived the rest-frame $V$-band luminosity function and its evolution up to $z=4$ using a galaxy sample selected in the near-infrared. For $z>4$, we use the UV luminosity functions of \cite{Bouwens2015}, at each redshift using the best-fit model flux in the band whose pivot wavelength, when transformed back to the rest-frame, is closest to $V$, and in which the off-centered emission has been detected. We exclude F070W and F090W to avoid potential effects from intergalactic medium absorption. Figure~\ref{p_overlap} shows the overlapping probability for the four targets. The final photometric redshift probability is obtained by multiplying the overlapping probability with the one derived from {\tt\string EAzY}, for all except the value at the spectroscopic redshift of the LRD point source, for if the off-centered emission is indeed at this redshift, physical correlation (Section~\ref{neb_model}) can exist between the blob and the LRD, in which case our assumption of a uniform galaxy spatial distribution is no longer applicable. We present the photometric redshift fitting results in Appendix~B, and further examine both the high- and low-redshift solutions under a more sophisticated stellar population synthesis framework in the following section.

\begin{figure*}[ht!] 
\centering
\includegraphics[width=\textwidth]{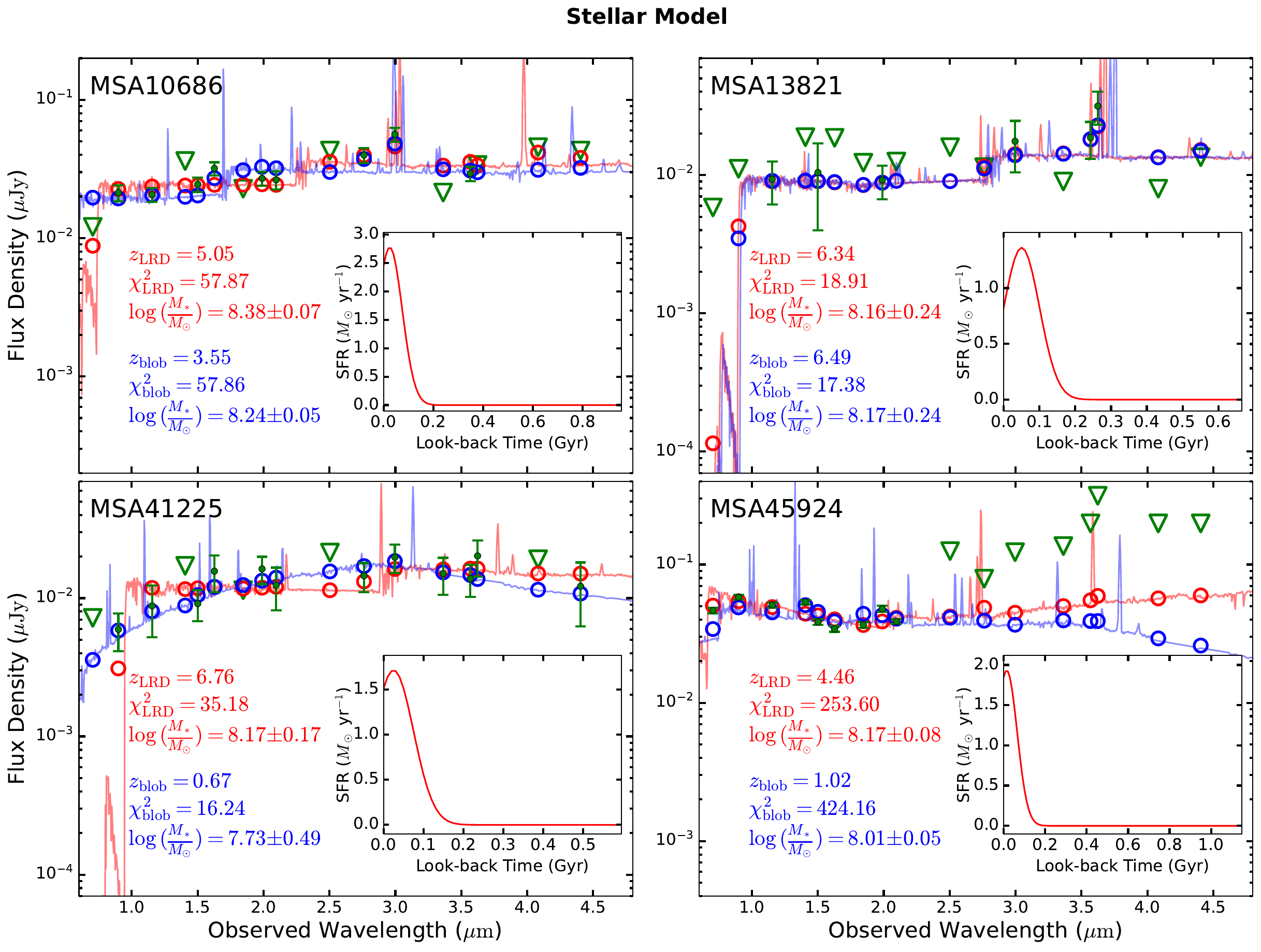}
\caption{Best-fit SED models from {\tt\string GalfitS} for the off-centered emission. The measured flux densities for solid detections are presented in filled green dots with error bars, and upper limits are marked as green triangles. Blue line and circles represent the best-fit model assuming the photometric redshift from {\tt\string EAzY}, while red line and circles denote the model that uses the spectroscopic redshift of the LRD. The redshift, total fitting $\chi^2$, as well as the best-fit stellar mass for each model are labeled in each panel. The best-fit star formation history (averaged in 10 Myr bins) for the model at the LRD redshift is plotted in the inset panel.
\label{gs_fitting_results}}
\end{figure*}

\begin{figure*}[ht!]
\centering
\includegraphics[width=\textwidth]{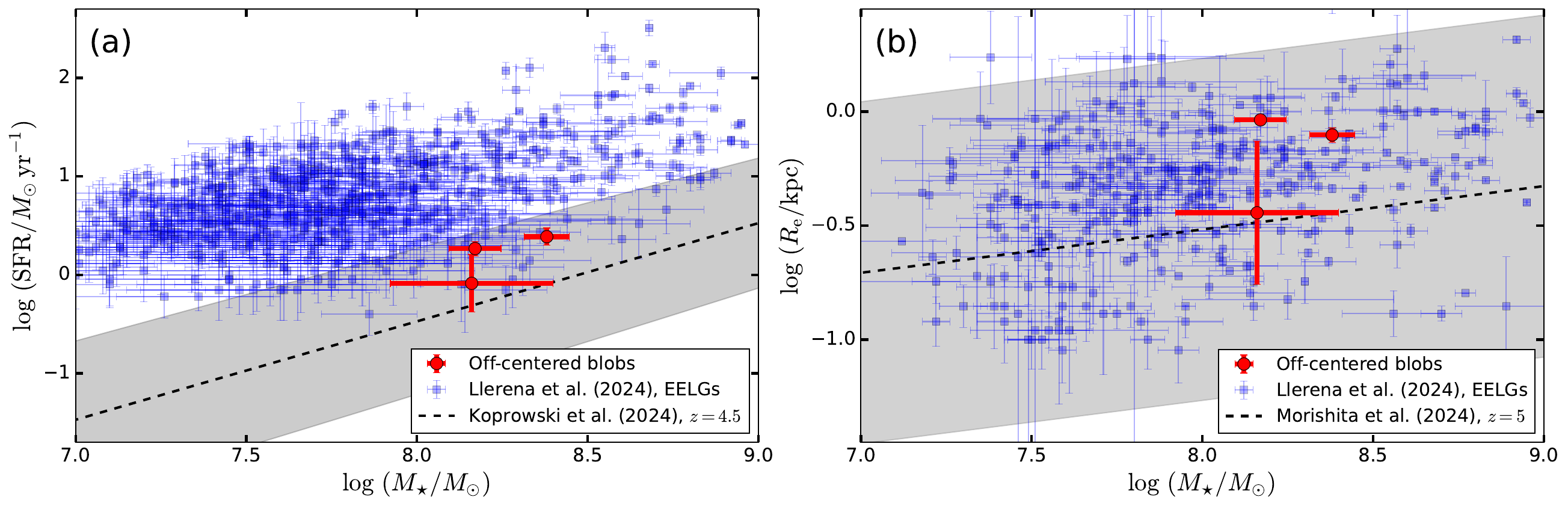}
\caption{The variation with stellar mass ($M_*$) with (a) star formation rate (SFR) and (b) stellar mass-effective radius ($R_e$) for the off-centered blobs (red dots) and high-redshift extreme emission-line galaxies (EELGs; blue dots) from \citet{Llerena2024}. The black dashed line in panel (a) represents the star formation main sequence at $z =4.5$ from \citet{Koprowski2024}, and in panel (b) the mass-size relation at $z=5$ from the evolution model of \citet{Morishita2024}. The grey-shaded region gives the $1\, \sigma$ scatter for the star formation main sequence in panel (a), and $3\, \sigma$ scatter for the mass-size relation in panel (b). We chose $3\, \sigma$ confidence region for the mass-size relation in order to be consistent with the comparison made in \citet{Llerena2024}.
\label{eelg_properties}}
\end{figure*}

\subsection{Stellar Model}
\label{galaxy_model}

Our photometric redshift analysis suggests that, apart from MSA13821, the physical association of the off-centered blobs and their nearby LRD is ambiguous: there exists two possible solutions, one similar to the spectroscopic redshift of the LRD point source and another at lower redshift. To evaluate further the robustness of these two solutions and to explore the physical properties of these potential companion galaxies, we model the observed multi-band photometry with the SED fitting module in {\tt\string GalfitS}. Specifically, model SEDs are calculated with the theoretical stellar templates from the Binary Population and Spectral Synthesis (BPASS V2.3; \citealt{Byrne2022}) code assuming a stellar initial mass function from \citet{Kroupa2001}. We adopt a continuous star formation history, which includes a stellar population undergoing continuous star formation since 200 Myr after the Big Bang, as well as a single stellar population formed in a burst that lasted for 100 Myr. In practice, the age of the single stellar population $t$ and the fraction of the total stellar mass formed during the burst ($f\_burst$) are treated as free parameters. Nebular emission from young stars is computed using the spectral synthesis code {\tt\string CLOUDY} \citep{Ferland1998, Chatzikos2023}, assigning the same gas-phase metallicity as the stellar metallicity. The combined templates are then corrected for dust extinction using the \citet{Calzetti2000} extinction curve and absorption from the intergalactic medium using the formalism of \citet{Meiksin2006} to produce the final galaxy SED model. Parameter ranges during the fitting are set as follows: $f\_burst$ between $10^{-2}$ to 1, metallicity $Z$ between 0.05 to 1\,$Z_\odot$, $A_V$ from 0 to 5 mag, and stellar mass $M_* = 10^{6.5}-10^{12}\,M_\odot$. We perform two fits, one at the spectroscopic redshift of the LRD point source and another at the best-fit photometric redshift. In each run, the age of the stellar population varies between $t = 0$ and the cosmic age minus 200 Myr at the corresponding redshift.

Figure~\ref{gs_fitting_results} presents the SED fits for both the high- and low-redshift solutions using stellar models. We slightly modify the manner in which we calculate the total $\chi^2$ of the fitting results. For the bands with solid detections, we sum up the residual value of $({\rm data}-{\rm model})^2/{\rm error}^2$ as in the conventional method, while for the bands with non-detections we only sum the residual values when the model fluxes exceed the upper limits. For MSA10686 and MSA13821, the low- and high-redshift solutions differ very little in terms of the total $\chi^2$ ($\Delta \chi^2<2$; $\Delta p\propto e^{\frac{\Delta \chi^2}{2}}<2.72$). By contrast, the low-redshift solution for MSA41225 has $\chi^2$ lower by 18.94, which translates to a factor of $\sim 1.3\times 10^4$ in terms of probability, while for MSA45924 the high-redshift solution, with $\Delta \chi^2=170.56$, gives a much better description to the data. Therefore, for all sources except MSA41225, considering the low overlapping probability (typically $\sim 10^{-6}-10^{-8}$), our results favor the high-redshift solutions consistent with the spectroscopic redshift of the LRD point sources over the low-redshift solutions given by {\tt\string EAzY}, firmly suggesting that the off-axis blobs are physically associated with the LRDs. In terms of MSA41225, we are less certain about its redshift due to the relatively large probability deficit of the high-z solution implied by the $\Delta \chi^2$. Therefore, in this and the following sections, we proceed under the assumption that the blobs are at same redshift as their LRD point sources, and restrict our analysis to MSA10686, MSA13821, and MSA45924, whose stellar models at the LRD redshift are equally good or show significant improvement compare to their low-redshift photometric redshift solutions. The best-fit parameters assuming the LRD redshift are reported in Table~\ref{par_sed}.

The [\ion{O}{3}] $\lambda\lambda 4959,\,5007$ and ${\rm H}\beta$ complex falls within the wavelength coverage of the medium-band filter F300M for MSA10686 and F360M for MSA13821. Both bands are detected unambiguously (Figure~\ref{gs_fitting_results}). The best-fit stellar SED model attributes the localized flux excess mostly to the [\ion{O}{3}] doublet, with an [\ion{O}{3}] $\lambda 5007$ equivalent width of 138\,\AA\ for MSA10686 and 256\,\AA\ for MSA13821. However, the SED models for these two targets underpredict the observed F300M flux by $22\%$ and the F360M flux by $40\%$. Assuming that all of the flux deficit originates from [\ion{O}{3}] and that the two lines have the same width and an intensity ratio of $\lambda 5007/\lambda 4959 = 2.98$ \citep{Storey2000}, the true equivalent width of [\ion{O}{3}] $\lambda 5007$ should be 276\,\AA\ for MSA10686 and 738\,\AA\ for MSA13821, akin to so-called extreme emission-line galaxies (EELGs; e.g., \citealt{vanderWel2011}). Figure~\ref{eelg_properties} compares the stellar mass, SFR, and effective radius of the off-centered blobs, within the context of the stellar model, with a sample of high-redshift EELGs at $4<z<9$ selected using empirical color-color criteria derived from templates based on local metal-poor starburst galaxies \citep{Llerena2024}. High-redshift EELGs lie systematically above the star-forming main sequence at $z=4.5$ \citep{Koprowski2024}, and they generally follow the stellar mass-size relation at $z=5$ \citep{Morishita2024}. The off-centered blobs, if they are indeed galaxies, have properties that are consistent with those of high-redshift EELGs.

\begin{deluxetable*}{cccccccc}[]
\label{par_sed}
\tabletypesize{\scriptsize}
\tablecaption{Best-fit SED Models}
\tablewidth{1pt}
\tablehead{ MSA  & \multicolumn{4}{c}{Stellar Model} & \multicolumn{3}{c}{Nebular Model} \\
                        \cmidrule(lr){2-5} \cmidrule(lr){6-8}
& $\log M_*$ & $\log {\rm SFR}$ & $t_{\rm age}$ & $A_V$ & $\log n_e$ & $Z$ & $f$ \\
& ($M_{\odot}$) & ($M_\odot\, {\rm yr}^{-1}$) & (Myr) & (mag) & ($\rm cm^{-3}$) & ($Z_{\odot}=0.02$) & \\
(1)&(2)&(3)&(4)&(5)&(6)&(7)&(8) }
\startdata
10686 & $8.38\pm0.07$ &   $0.39\pm0.088$ & $25.10\pm5.10$ & $0.18\pm0.060$   & $1.00\pm0.45$ & $0.063\pm0.014$ &  $3.38\pm0.14$ \\
13821 & $8.16\pm0.24$ & $-0.084\pm0.29$  & $50.10\pm13.90$ & $0.00060\pm0.079$ & $1.22\pm0.94$ & $0.094\pm0.024$ &  $5.08\pm0.92$ \\
45924 & $8.17\pm0.08$ &   $0.27\pm0.073$ & $14.50\pm0.70$ & $0.0043\pm0.022$ & $1.00\pm1.89$ & $0.0050\pm0.023$ &  $6.21\pm0.66$ \\
\enddata
\tablecomments{Col. (1): Name. Col. (2): Total stellar mass. Col.(3): Star formation rate averaged over the last 10 Myr. Col. (4): Age of the recent starburst. Col. (5): Dust extinction in the $V$ band. Col. (6): Nebular electron density. Col. (7): Nebular gas metallicity. Col. (8): Scaling factor of the nebular template. }
\end{deluxetable*}

\begin{figure*}[ht!]
\centering
\includegraphics[width=\textwidth]{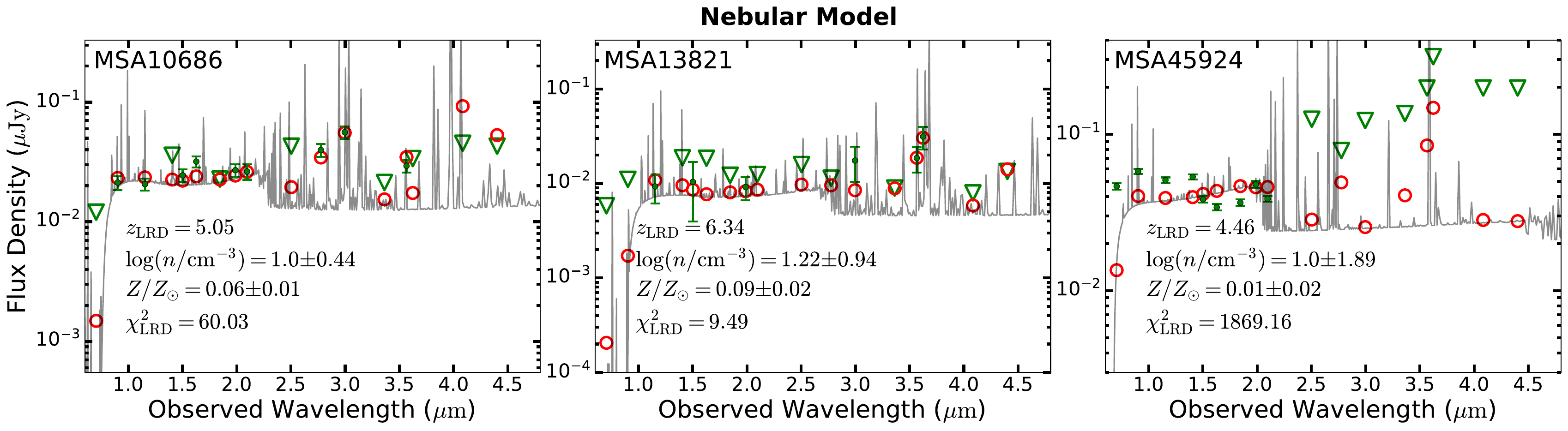}
\caption{Same as Figure~\ref{gs_fitting_results} but for the nebular models generated with {\tt\string CLOUDY}. Redshift as well as best-fit density and metallicity are given in each panel. 
\label{nebular_fitting_results}}
\end{figure*}

\subsection{Nebular Model}
\label{neb_model}

While the stellar model can generally reproduce the observed SED, it underpredicts the fluxes from the [\ion{O}{3}]$+\rm H\beta$ complex, indicating a potentially different, nonstellar source of ionization. We explore the possibility that the off-centered emission originates from gas photoionized by the radiation from the nearby LRD point source. We calculate a grid of spectral templates using {\tt\string CLOUDY}, adopting an open geometry with the inner and stop radius set to $r_{0,{\rm c}}-r_{{\rm e,c}}$ and $r_{0,{\rm c}}+r_{{\rm e,c}}$, respectively, with $r_{{\rm e,c}}$ the effective radius of the off-centered emission and $r_{0,{\rm c}}$ the distance from the LRD point source. The output spectrum is especially sensitive to the shape of the ionizing spectrum in the extreme-UV, which contributes the majority of the ionizing photon budget. We construct the input ionizing spectrum by interpolating the observed multi-band photometry of the LRD point source. As we cannot directly measure the continuum shorter than the Lyman limit due to intergalactic medium absorption, we assume that it follows the same power law redder than the Lyman limit\footnote{We have verified that under this approximation and case B recombination the number of Lyman continuum photons suffices to produce the observed ${\rm H}\alpha$ luminosities reported in \citet{Greene2024}.}. We calculate the spectral slopes by fitting a power-law function to all the detected broad-band photometry of the LRD point sources whose pivot wavelength is shorter than 3500\,\AA, as the spectral turnover for LRDs generally occurs near the Balmer limit. Only broad-band photometry is used to avoid possible contamination of emission lines to the medium bands, but we exclude F070W and F090W because they are affected by intergalactic medium absorption. Expressing the UV power law continuum as $f_{\nu}\propto \, {\nu}^\alpha$, $\alpha = -0.77$, $-0.91$, $-0.41$, and $-2.12$ for MSA10686, MSA13821, MSA41225, and MSA45924, respectively. The fraction of the ionizing radiation that enters the blobs is calculated using a covering factor estimated from the solid angle of a sphere centered on the LRD whose radius and position equal the effective radius and the center of the off-centered blob. We compute spectra with different combinations of the density and metallicity, with ${\rm log}\,(n/{\rm cm}^{-3}) = 1-4$ in steps of 1~dex and $Z = 0.005$, 0.02, 0.2, 0.4, and $1\,Z_\odot$ assuming solar elemental abundances. We then correct the output spectra for intergalactic medium absorption following the formalism \citep{Meiksin2006} used for our stellar model, before interpolating them as a two-variable function of $\log n$ and $Z$ using the Python package {\tt\string Scipy.interpolate.RegularGridInterpolator} to obtain the final nebular spectral models. Least-squares fitting is performed to search for the nebular parameters that produce a spectrum bearing the closest resemblance to the observed SEDs. The fitting includes a free parameter $f$, which varies between 1 and 10, to scale the total flux of the model spectrum, to account for potential effects from non-spherical geometry as well as inhomogeneity in density and metallicity distribution, which are not considered in our {\tt\string CLOUDY} calculations. 

The fitting results for the nebular model are shown in Figure~\ref{nebular_fitting_results} and reported in Table~\ref{par_sed}. Overall, the nebular model describes well the SEDs of MSA10686 and MSA13821, especially in the band with a flux upturn due to strong line emission from [\ion{O}{3}] and ${\rm H}\beta$. For MSA13821, the nebular model achieves a better fit ($\Delta \chi^2=9.42$) compared to the stellar model. The result from the nebular model of MSA10686 has similar goodness-of-fit as the stellar model, with slightly higher $\chi^2$ ($\Delta \chi^2=2.16$). The best-fit models for MSA10686 and MSA13821 suggest the off-center, extended blob arises from low-density ($n \approx 10\,{\rm cm}^{-3}$), metal-poor ($Z\approx 0.05-0.1\,Z_\odot$) gas. Note that while the best-fit density of MSA10686 reaches the lower limit of our parameter grid, we are unable to explore results of lower density, as it causes catastrophic failure in {\tt\string CLOUDY} due to large negative optical depth in line transition.

By contrast, the nebular model failed to fit the SED of MSA45924. Most of the mismatch between data and model occurs in the bluest end of the SED sampled by F070W and F090W, where the turnover in the nebular spectrum due to two-photon decay underpredicts the observed flux.

\section{Discussion}\label{discussion}

\subsection{Nature of the Off-centered Blobs}

Off-centered, extended emission is associated with a significant fraction of LRDs. In their systematic study of 99 photometrically selected LRDs at $z \approx 4-8$, \cite{Rinaldi2025} find that $\sim 30\%$ show a variety of complex, asymmetric UV morphologies beyond the dominant central point source. This is consistent with the results of \cite{Chen2025}, based on much meager statistics.  What is the physical nature of this asymmetric, resolved component? Assuming that the off-centered emission near LRDs is of stellar origin, we show that their inferred SFRs, stellar masses, and effective radii qualitatively resemble those of high-redshift EELGs. Under this interpretation, the off-centered blobs may be regarded as companion galaxies, which are presumably interacting or merging with a large fraction of the LRDs. While LRDs lack a galaxy component \citep{Chen2025} massive enough to be commensurate with the local scaling relations between black hole and galaxy mass \citep{Kormendy2013, Greene2020}, gas tidally stripped from these star-forming galaxies could be the fueling channel that maintains these black holes in an active state. Alternatively, if the off-centered blobs comprise stars of the LRD host galaxies themselves, then their global asymmetry can arise from dynamical processes that can dislodge the black hole to kpc-scale distances, such as gravitational recoil \citep{Bekenstein1973, Campanelli2007} or gravitational slingshot effects \citep{Hoffman2007}, indicating that the central black holes in the LRDs are produced by merging from multiple lower mass black holes. The stellar scenario offers the best interpretation for the off-centered emission in MSA45924.

Extended structures observed in broad-band images, especially near luminous AGNs like the LRDs, are not necessarily of stellar origin. We explore an alternative possibility, in which the off-centered blobs constitute nebular gas directly ionized by the radiation field emanating from the adjacent LRD. Our photoionization calculations show that this scenario is viable, for two of the four sources studied here. Using the observed SED of the LRD as the ionizing radiation source and the relative displacement between the off-centered component and the point source, the nebular model can very well explain the observed flux and SED shape of MSA10686 and MSA13821, even better than the stellar model in the case for MSA13821. The nebular model has notable success in accounting for the high equivalent width of the [\ion{O}{3}] line detected in the medium-band filter F300M for MSA10686 and F360M for MSA13821. This is reminiscent of the situation in high-redshift galaxies whose hard ionizing radiation field produces sufficient nebular emission to affect the broad-band flux \citep[e.g.,][]{Cameron2024,Eugenio2025, Kataria2025, Solimano2025}. For these two targets, the best-fit nebular model gives a density of $\sim 10\, {\rm cm}^{-3}$, with radius-averaged electron temperature of $\sim 10^4\, {\rm K}$. Since the best-fit density already reaches the lowest value of our photoionization calculations, the real density could be lower. The low density and low surface brightness of the nebular emission qualitatively resembles the warm ionized medium ubiquitously detected in local spiral galaxies (e.g., \citealt{Wang1998}). Interestingly, although under the nebular scenario the extended emission does not require necessarily the presence of any galaxy, the best-fit metallicity for MSA10686 and MSA13821 are both $Z \approx 0.05-0.1\,Z_{\odot}$, much higher than that of the intergalactic medium at cosmic noon. According to \citet{Simcoe2004} and \citet{McQuinn2016}, only $\lesssim 10\%$ of sight lines with \ion{O}{4} or \ion{C}{4} absorption at $z\approx 2.5$ have metallicities higher than $0.01\,Z_{\odot}$. At redshifts closer to the sources in our sample, the 95\% upper limit of the intergalactic medium [Mg/H] ratio, defined as the logarithmic number ratio of magnesium atoms to hydrogen atoms relative to the solar ratio, is $-3.73$ at $z = 6.47$ \citep{Tie2024}, suggesting $Z<10^{-3}Z_{\odot}$. The nebular gas could have been chemically enriched by stars in a preexisting, albeit undermassive \citep{Chen2025}, host galaxy. or by star formation connected with the accretion disk \citep[e.g.,][]{Artymowics1993, Zhang2025}.

Future JWST observations with spatially resolved spectra, such as those that can be obtained with the integral-field units of NIRSpec or via slitless spectroscopy with NIRCam or NIRISS, can provide more definitive diagnostics of the physical nature of the extended, off-centered emission associated with LRDs. For example, our nebular model for MSA1068 predicts prominent [\ion{S}{2}] $\lambda\lambda 6716,\,6731$ emission, similar to the situation seen in the warm ionized medium in the local Universe that shares similar density and temperature as our best-fit model spectrum. Detection of [\ion{S}{2}] and measurement of the [\ion{S}{2}]/${\rm H}\alpha$ ratio for the extended emission would provide smoking gun support for their nebular nature.

\subsection{Estimation of Gas Mass and Halo Accretion Rate}

We estimate the total mass of ionized gas in both models. Under the stellar scenario, in which the continuum is assumed to be produced by stars, photometry in F300M and F360M provides constraints on the total flux of the [\ion{O}{3}]$+{\rm H}\beta$ complex in MSA10686 and MSA13821, respectively. We convert this value to ${\rm H}\beta$ luminosity using the relation between [\ion{O}{3}]/${\rm H}\beta$ ratio and specific star formation rate derived in \citet{Dickey2016}. The gas mass is then calculated using the dereddened ${\rm H}\beta$ luminosity, assuming $n_{e}=100\,{\rm cm}^{-3}$, $T=10^4\, {\rm K}$, and an effective recombination coefficient of $\alpha_{{\rm eff},{\rm H}\beta}=3.03\times 10^{-14}\, {\rm cm}^{-3}$ \citep{Draine2011},

\begin{figure}[t]
\centering
\includegraphics[width=0.49\textwidth]{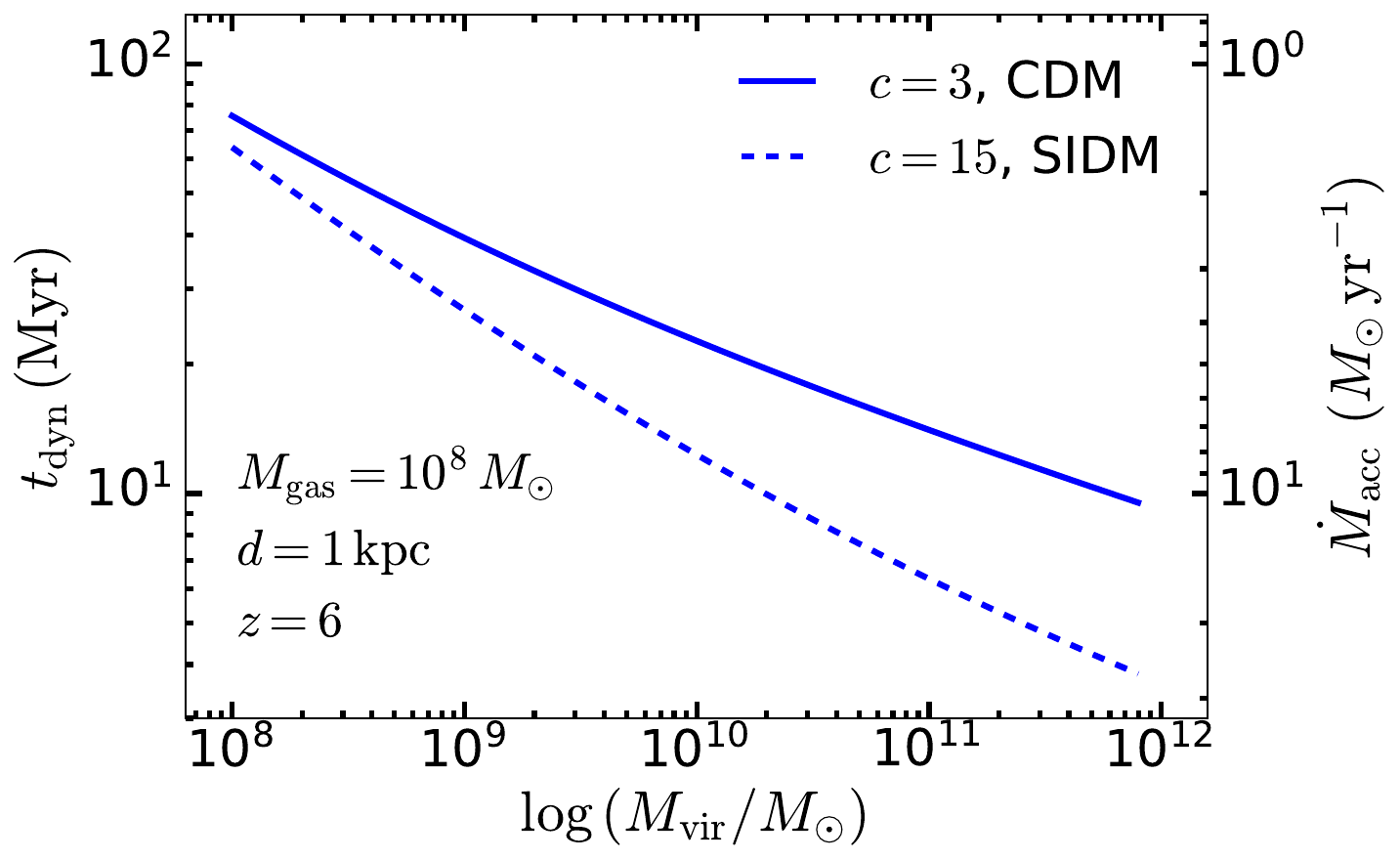}
\caption{Dynamical timescale (left vertical axis) and the corresponding baryonic mass accretion rate (right vertical axis) as functions of halo virial mass at $z=6$, for a $10^8\, M_{\odot}$ gas cloud infalling from 1 kpc. The blue solid (dashed) line represents result for halo concentration $c=3$ ($c=15$), respectively.
\label{t_ff_M_dot}}
\end{figure}

\begin{equation}
M_{\rm gas}=1.15\times 10^4\,\left(\frac{n_{e}}{100\,{\rm cm}^{-3}}\right)^{-1}\,\left(\frac{L_{{\rm H}\beta}}{10^{38}\,{\rm erg}\,{\rm s}^{-1}}\right)\, M_{\odot}
\end{equation}

\noindent

If the off-centered blobs near MSA10686 and MSA13821 are galaxies, their estimated ionized gas mass is ${\rm log}\,(M_{\rm gas}/M_{\odot})=6.96\pm0.05$ and $6.93\pm0.12$, respectively\footnote{We bootstrap all parameter uncertainties through each step of the calculation and take the final standard deviation as the error.}. Under the nebular scenario, the broad-band emission is attributed to the recombination continuum. In this case, we estimate the total ionized gas mass using the best-fit electron density and total emission volume of our {\tt\string CLOUDY} model, multiplied by the best-fit scaling factor $f$. For MSA10686 and MSA13821, their best-fit nebular templates correspond to ${\rm log}\,(M_{\rm gas}/M_{\odot})=8.71\pm0.28$ and $7.90\pm0.54$, higher than the prediction from the stellar model by more than an order of magnitude.

If the off-centered gas clouds on kpc scales are infalling within the halo of the LRDs, their gas mass provides a rough estimate of the LRD halo mass accretion rate. The infall time of the gas clouds onto the LRD halo can be estimated as 

\begin{equation}
t_{\rm dyn}=1.65\,{\rm Myr}\, \left(\frac{d}{1\,{\rm kpc}} \right)^{3/2}\, \left(\frac{M_{h,\, {\rm cen}}}{10^{11}\,M_{\odot}} \right)^{-1/2},
\end{equation}

\noindent
where $d$ is the distance between the off-centered blobs and the LRD point source, $M_{h,\, {\rm cen}}$ is the central enclosed halo mass within radius $d$, which is a function of both the virial mass of the entire halo ($M_{\rm vir}$), the halo concentration $c$ and redshift $z$. The baryonic mass accretion rate is therefore

\begin{equation}
\dot{M}_{\rm acc}\sim \frac{M_{\rm gas}}{t_{\rm dyn}} = 100\, M_{\odot}\,{\rm yr}^{-1}\, \left(\frac{M_{\rm gas}}{10^8\, M_{\odot}} \right)\, \left(\frac{t_{\rm dyn}}{1\,{\rm Myr}} \right)^{-1}
\end{equation}

\noindent

These estimates depends on our highly incomplete knowledge of the halo masses of LRDs. 
Clustering analysis of 27 low-luminosity broad-line AGNs discovered by JWST at $5 < z < 6$, among them LRDs, constrain their typical halo mass to $M_h \approx 3\times 10^{11}\, M_{\odot}$ \citep{Arita2025}. 
On the other hand, \cite{Jiang2025} recently proposed a model for the formation of seed black holes in LRDs via the gravothermal collapse of the cores of self-interacting dark matter (SIDM) halos. Within this framework, core collapse occurs most efficiently within a relatively narrow range of halos with masses $M_h \approx 10^8-10^{10}\,M_{\odot}$, significantly lower than that suggested by the clustering analysis of \cite{Arita2025}. In Figure~\ref{t_ff_M_dot}, we present the value of $t_{\rm dyn}$ and $\dot{M}_{\rm acc}$ as functions of $M_{\rm vir}$ for halos with $c=3$, which is typical for the conventional structure formation framework assuming cold dark matter (CDM), and halos with $c=15$ as proposed by \cite{Jiang2025} in the SIDM scenario. 

For typical CDM halos with $c=3$ at $z=6$, if the $3\times 10^{11}\, M_{\odot}$ halo mass estimated by \citet{Arita2025} can be applied to typical LRDs, the infall time of a $10^8\,M_{\odot}$ gas cloud starting at 1 kpc scale is $t_{\rm dyn} \approx 11\,{\rm Myr}$, and the accretion rate $\dot{M}_{\rm acc} \approx 9\, M_{\odot}\,{\rm yr}^{-1}$. Alternatively, adopting the lower halo mass ($10^8-10^{10}\,M_{\odot}$) predicted for the SIDM halos with $c=15$, the infall time increases to $t_{\rm dyn} \approx 12-64\,{\rm Myr}$, and the accretion rate drops to $\dot{M}_{\rm acc} \approx 2-8\, M_{\odot}\,{\rm yr}^{-1}$.
For an LRD at $z=6$, if the halo accretion rate is sustained to $z=4$, the accreted gas would form a galaxy with a total stellar mass $M_* \approx 10^9\,M_{\odot}$ assuming a star formation efficiency of 0.1. The stellar mass would be even higher if stars form more efficiently at high redshift (e.g., \citealt{Inayoshi2022, Ceverino2024}). The large stellar mass assembled would gradually render the LRD spatially resolved, which explains the rapid decline in the number density of compact sources with the characteristic ``V-shape'' SEDs at $z<4$ reported by both observations \citep{Kocevski2025, Ma2025} and theoretical predictions \citep{Inayoshi2025a}.

\section{Summary}\label{summary}

A significant fraction of LRDs exhibits nearby extended emission of unknown origin. If physically associated with the LRD, this component may trace stellar emission from an off-centered host galaxy or neighboring companions, or nebular gas illuminated by the active nucleus. We perform simultaneous 17-band morphological decomposition for four LRDs in the UNCOVER and MegaScience surveys, and for the first time report the continuum shape and properties of the [\ion{O}{3}]$+\rm H\beta$ complex for the off-centered extended emission that are commonly detected in the immediate vicinity of LRDs. We evaluate whether these emission blobs are physically associated with the LRDs by estimating their photometric redshift and probability of foreground galaxy overlapping. These measurement enable us to further understand the physical nature of these blobs by analyzing the observed SEDs with both a stellar model and pure nebular emission model. Our main conclusions are as follows:

\begin{itemize}
\item Combining the photometric redshift and overlapping probability, we confirm that three of our targets (MSA10686, MSA13821, MSA45924) are at a similar redshift as the LRDs, while the fourth (MSA41225) is more consistent with a projected galaxy of lower redshift.

\item Fitting the SEDs of the extended emission physically associated with the remaining three LRDs, the stellar model yields star formation rates, stellar masses, and effective radii that are consistent with the properties of high-redshift extreme emission-line galaxies. However, only the blob of MSA45924 is best fit by the stellar model. The extended emission associated with MSA13821, and to a lesser extent also that of MSA10686, is consistent with nebular emission from low-density ($n<10\, {\rm cm}^{-3}$), low-metallicity ($Z\approx 0.05-0.1\,Z_{\odot}$) gas photoionized by the ultraviolet radiation from the nearby LRD.

\item Using current constraints on the halo masses of the LRDs estimated from clustering analysis and theoretical considerations, we estimate a typical baryonic halo gas mass accretion rate of $\sim 2-9 \, M_{\odot}\,{\rm yr}^{-1}$, which, if unabiated and converted to stars until $z = 4$ with an efficiency of 10\%, would build sufficient stellar mass to reveal the extended host galaxy of the LRD at lower redshifts.

\end{itemize}

\section*{Acknowledgments}
\begin{acknowledgments}
This work was supported by National Key R\&D Program of China (2022YFF0503401), the China Manned Space Program (CMS-CSST-2025-A09), and the National Science Foundation of China (12233001). CHC thanks Fangzhou Jiang, Zhengrong Li, Claudio Ricci, and Zijian Zhang for helpful comments and discussions.
\end{acknowledgments}

\appendix

\section{Image Fitting Results}

\begin{figure*}[h] 
\centering
\figurenum{A1}
\includegraphics[width=0.49\textwidth]{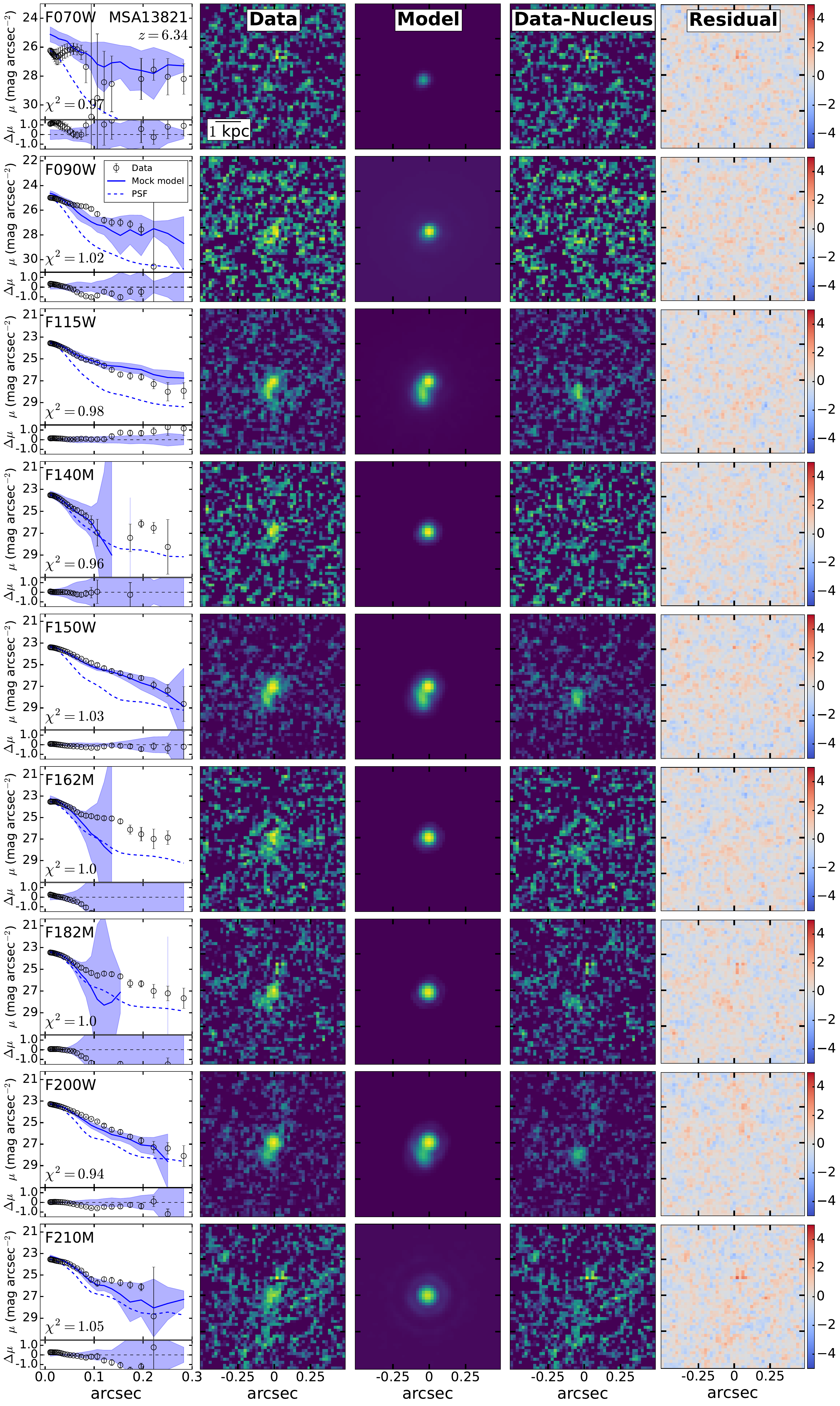}
\includegraphics[width=0.49\textwidth]{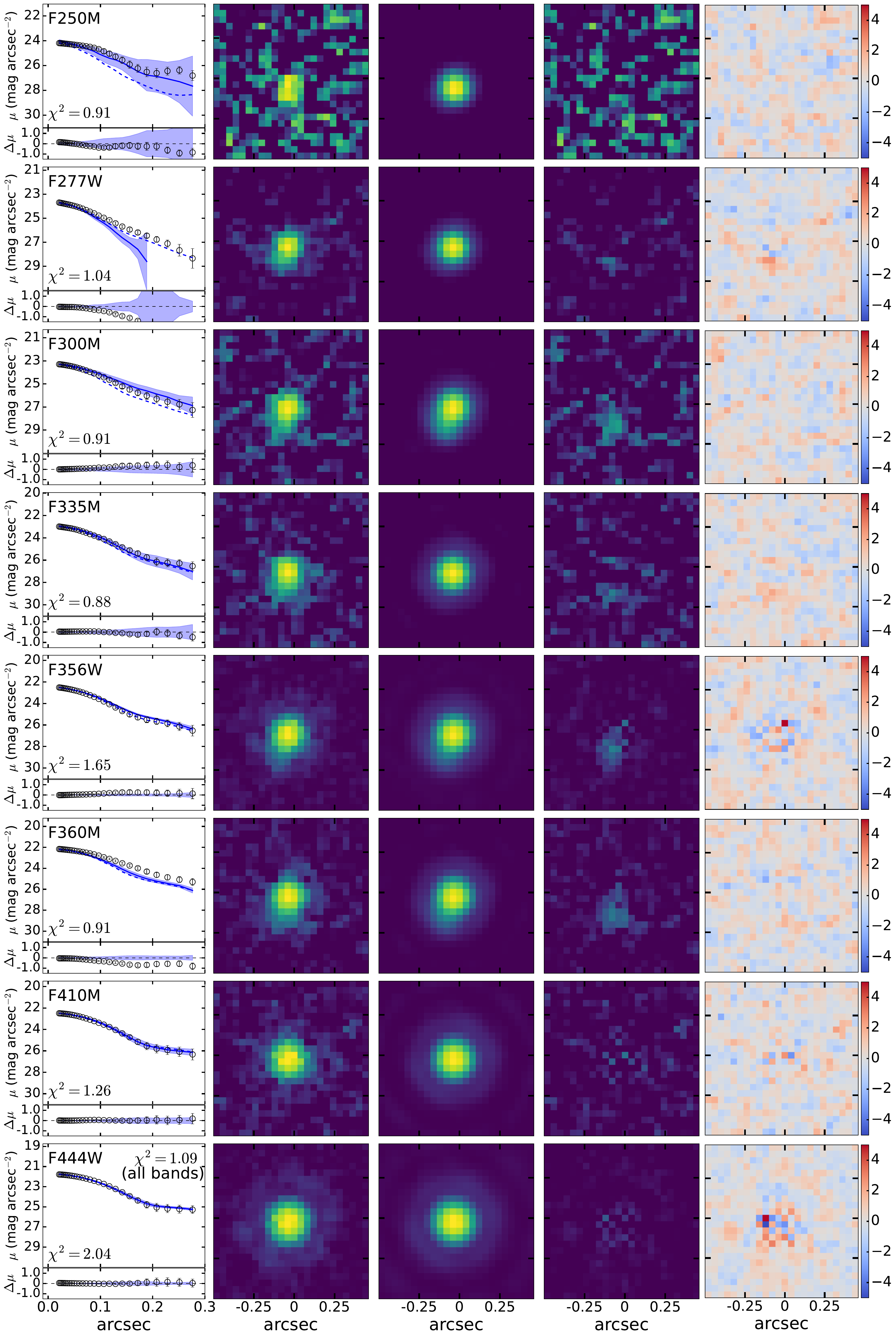}
\caption{Same as Figure~\ref{10686_fitting_results}, but for MSA13821.}
\label{13821_fitting_results}
\end{figure*}

\begin{figure*}[h]
\centering
\figurenum{A2}
\includegraphics[width=0.49\textwidth]{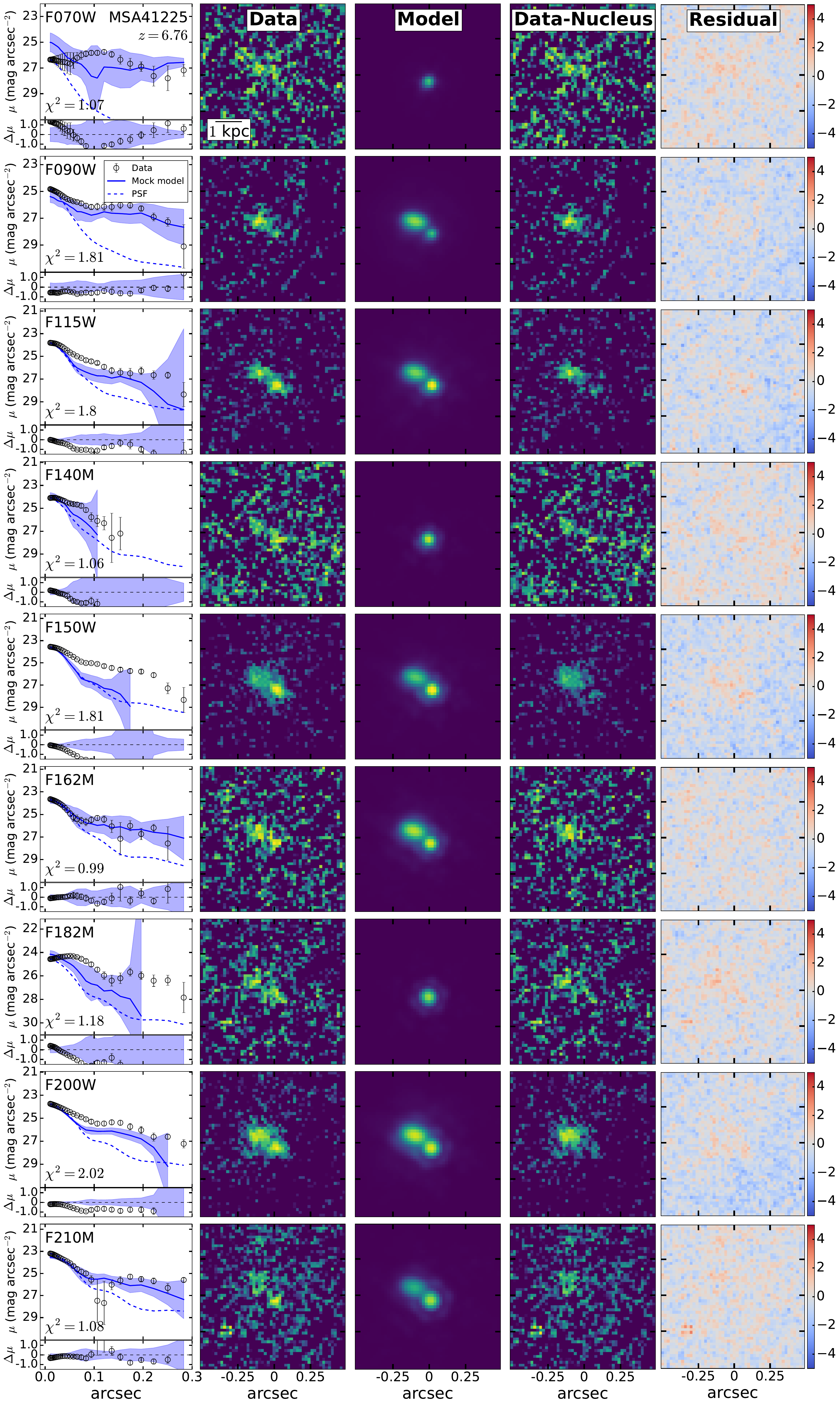}
\includegraphics[width=0.49\textwidth]{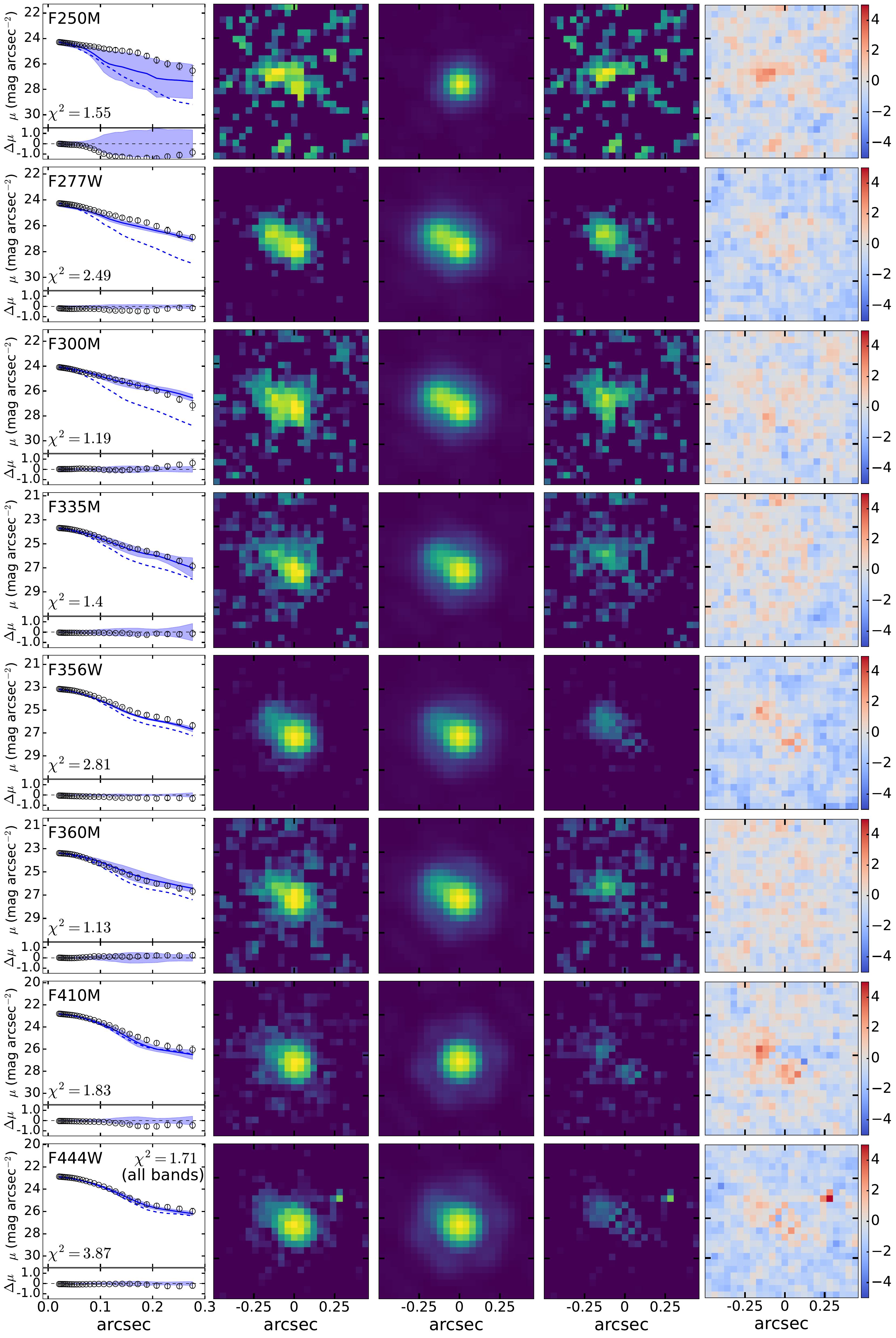}
\caption{Same as Figure~\ref{10686_fitting_results}, but for MSA41225.}
\label{41225_fitting_results}
\end{figure*}

\begin{figure*}[h]
\centering
\figurenum{A3}
\includegraphics[width=0.49\textwidth]{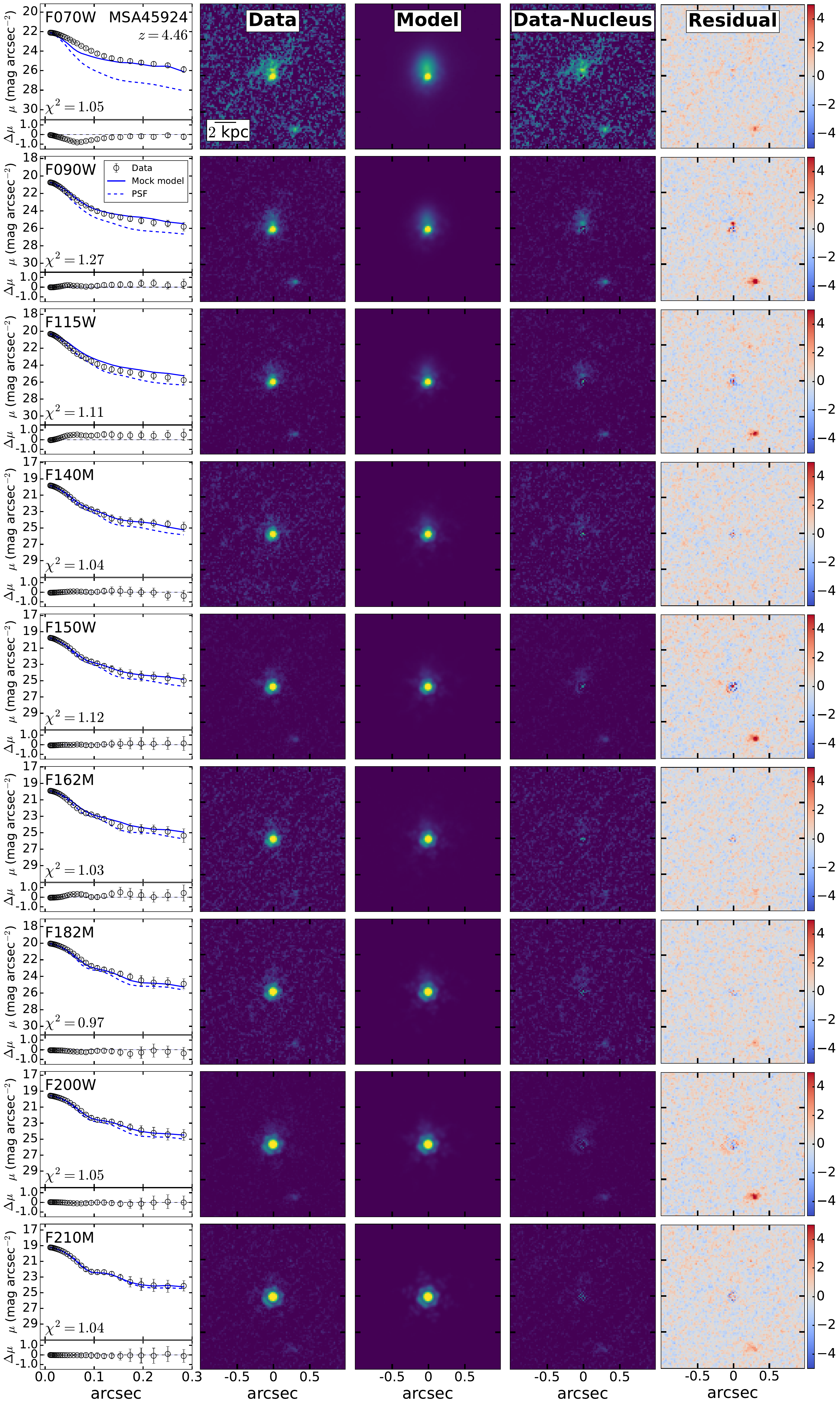}
\includegraphics[width=0.49\textwidth]{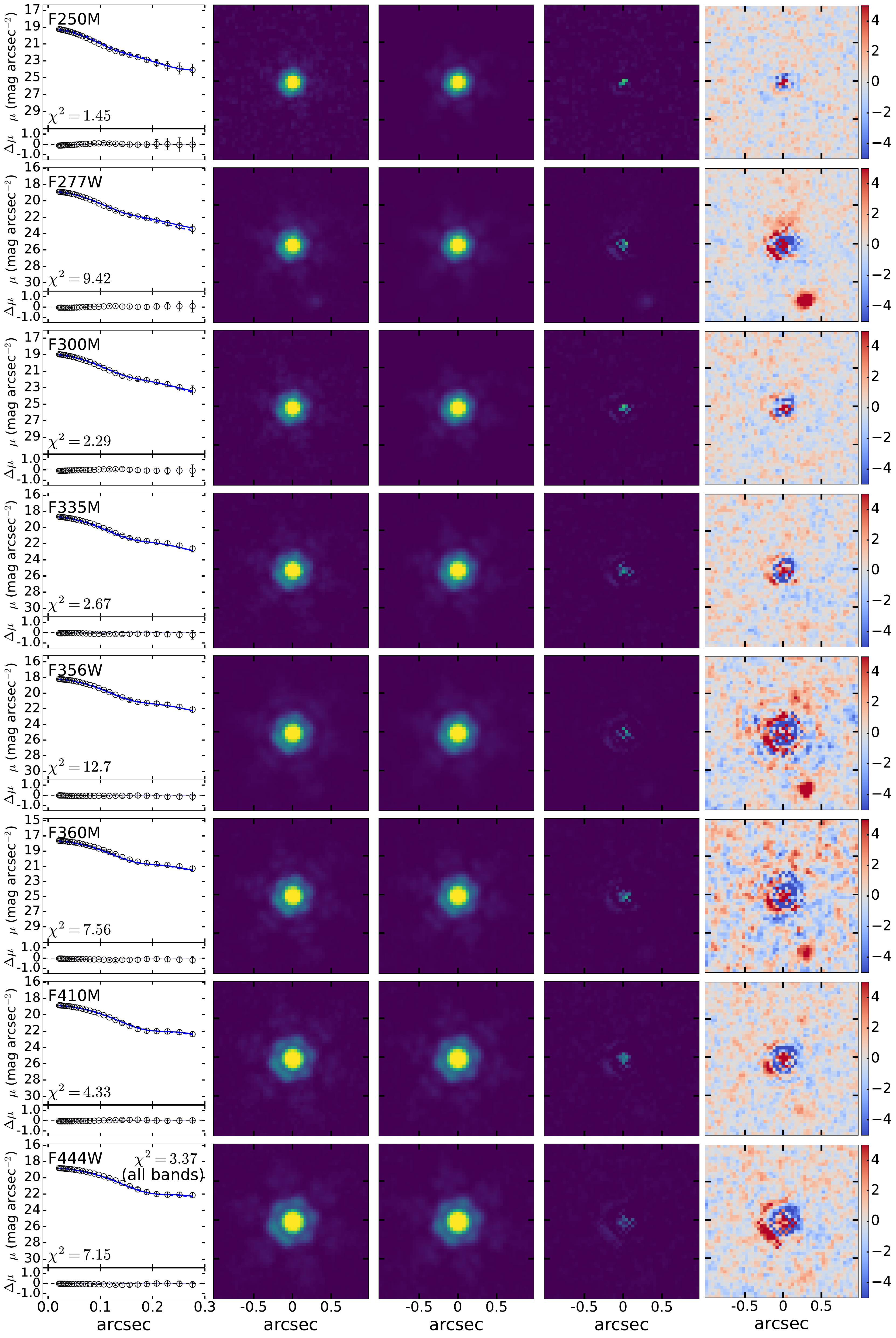} 
\caption{Same as Figure~\ref{10686_fitting_results}, but for MSA45924.}
\label{45924_fitting_results}
\end{figure*}

The fitting results for MSA13821, MSA41225, and MSA45924 are shown in Figures~\ref{13821_fitting_results}, \ref{41225_fitting_results}, and \ref{45924_fitting_results}, respectively.

\newpage

\newpage

\newpage

\section{Photometric Redshift Fitting Results}

\begin{figure*}[h]
	\centering
        \figurenum{B1}
	\includegraphics[width=\textwidth]{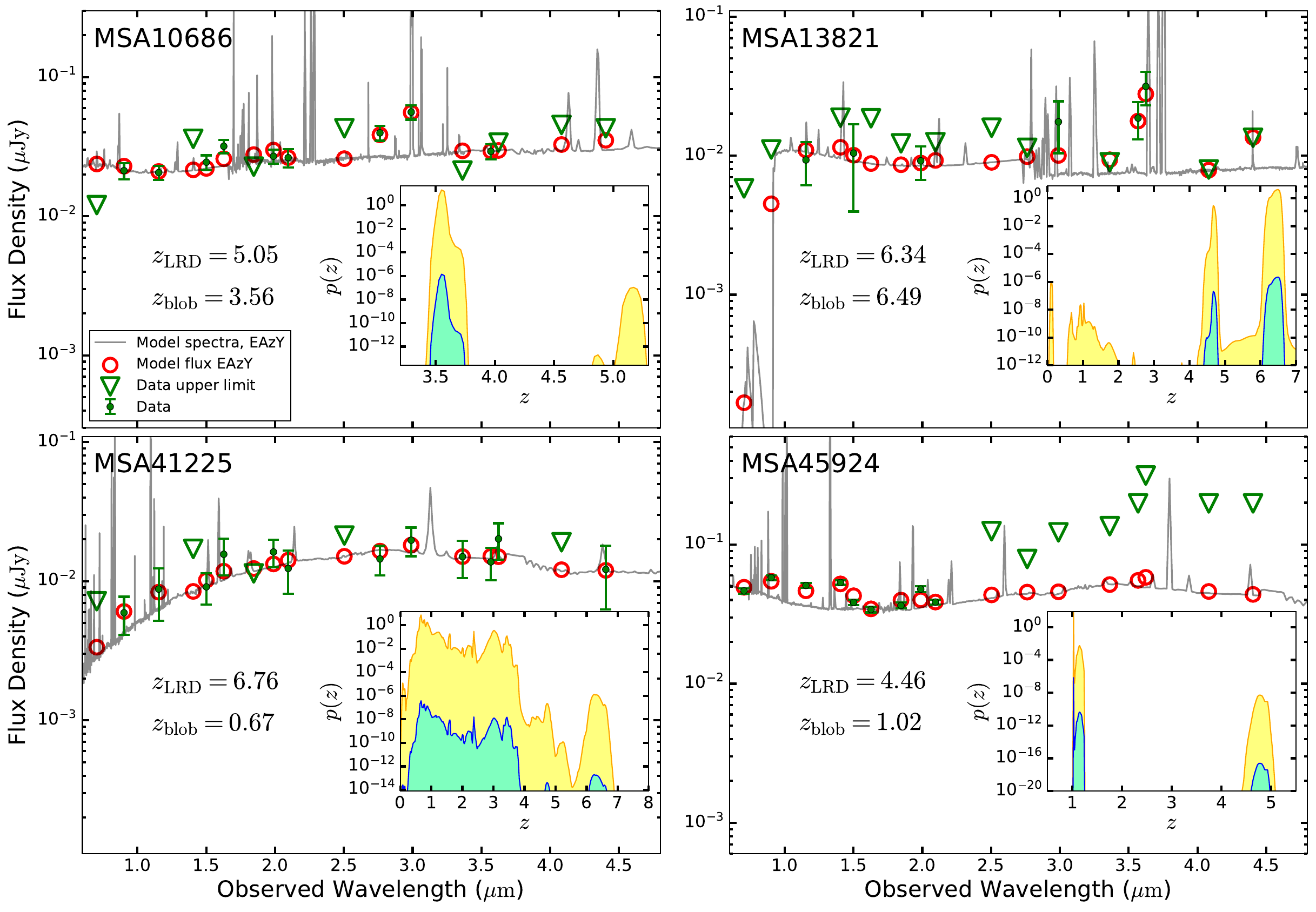}
	\caption{Best-fit SED models from {\tt\string EAzY} for the off-centered emission. The measured flux densities for solid detections are presented in filled green dots with error bars, while upper limits are 		marked as green triangles. The spectroscopic redshift for the LRD point source ($z_{\rm LRD}$) as well as the redshift of the off-centered blob with the highest probability ($z_{\rm blob}$) is given in each 	panel. The inset panel shows the probability distribution function [$p(z)$] of redshift, with the orange shaded regions represent the original probability derived by {\tt\string EAzY}, and the cyan shaded regions represent the probability rescaled by the overlapping probability that the off-centered blobs correspond to galaxies at redshifts different from the LRD point sources. 
	\label{eazy_fitting_results}}
	\end{figure*}

We give the results for photometric redshift fitting with {\tt\string EAzY} in Figure~\ref{eazy_fitting_results}. The best-fit templates, probability distribution of the redshift solutions before and after multiplying the overlapping probability are both given in each panels for all the sources. Detailed procedures are described in Section~\ref{phot-z}.

\end{CJK*}

\begin{thebibliography}{}

\bibitem[Akins {et~al.}(2025)]{Akins2025} Akins, H., Casey, C.~M., Lambrides, E., {et~al.} 2025, \apj, submitted (arXiv:2406.10341)

\bibitem[Arita {et~al.}(2025)]{Arita2025} Arita, J., Onoue, M., Yoshioka, T., {et~al.} 2025, \mnras, 536, 3677

\bibitem[Artymowics {et~al.}(1993)]{Artymowics1993} Artymowics, P., Lin, D.~N.~C., Wampler, E.~J. 1993, \apj, 409, 592

\bibitem[Baggen {et~al.}(2024)]{Baggen2024} Baggen, J.~F.~W., van Dokkum, P., Brammer, G., {et~al.} 2024, \apj, 977, L13

\bibitem[Bekenstein(1973)]{Bekenstein1973} Bekenstein, J.~D. 1973, \apj, 183, 657

\bibitem[Bennert {et~al.}(2006)]{Bennert2006a} Bennert, N., Jungwiert, B., Komossa, S., {et~al.} 2006, A\&A, 456, 953

\bibitem[Berger(2010)]{Berger2010} Berger, E. 2010, \apj, 722, 1946

\bibitem[Bertin(2013)]{Bertin2013} Bertin E. 2013, PSFEx: Point Spread Function Extractor, Astrophysics Source Code Library, record ascl:1301.001

\bibitem[Bertin \& Arnouts(1996)]{Bertin1996} Bertin, E., \& Arnouts, S. 1996, A\&AS, 117, 393

\bibitem[Bezanson {et~al.}(2024)]{Bezanson2024} Bezanson, R., Labb\'e, I., Whitaker, K.~E., {et~al.} 2024, ApJ, 974, 92

\bibitem[Bloom {et~al.}(2002)]{Bloom2002} Bloom, J.~S., Kulkarni, S.~R., Djorgovski, S.~G. 2002, \aj, 123, 1111

\bibitem[Bouwens {et~al.}(2015)]{Bouwens2015} Bouwens, R.~J., Illingworth, G.~D., Oesch, P.~A., {et~al.} 2015, \aj, 803, 34


\bibitem[Brammer {et~al.}(2022)]{Brammer2022} Brammer, G., Strait, V., Matharu, J., {et~al.} 2022, {grizli}, 1.5.0,  Zenodo

\bibitem[Brammer {et~al.}(2008)]{Brammer2008} Brammer, G., van Dokkum, P.~G., Coppi, P. 2008, \apj, 686, 1503

\bibitem[Byrne {et~al.}(2022)]{Byrne2022} Byrne, C.~M., Stanway, E.~R., Eldridge, J.~J., {et~al.} 2022, \mnras, 512, 5329

\bibitem[Calzetti {et~al.}(2000)]{Calzetti2000} Calzetti, D., Armus, L., Bohlin, R.~C., {et~al.} 2000, \apj, 533, 682

\bibitem[Cameron {et~al.}(2024)]{Cameron2024} Cameron, A.~J., Katz, H., Witten, C., {et~al.}, \mnras, 534, 523

\bibitem[Campanelli {et~al.}(2007)]{Campanelli2007} Campanelli, M., Lousto, C., Zlochower, Y., {et~al.} 2007, \apj, 659, L5

\bibitem[Ceverino {et~al.}(2024)]{Ceverino2024} Ceverino, D., Nakazato, Y., Yoshida, N., Klessen, R., \& Glover, R. 2024, \aap, 689, A244

\bibitem[Chatzikos {et~al.}(2023)]{Chatzikos2023} Chatzikos, M., Bianchi, S., Camilloni, F., {et~al.}, 2023, RMxAA, 59, 327

\bibitem[Chen {et~al.}(2025)]{Chen2025} Chen, C.-H., Ho, L.~C., \& Li, R., 2025, \apj, 983, 60

\bibitem[Conroy {et~al.}(2010)]{Conroy2010} Conroy, C., \& Gunn, J.~E. 2010, \apj, 712, 833

\bibitem[D'Eugenio {et~al.}(2025)]{Eugenio2025} D'Eugenio, F., Helton, J.~M., Hainline, K., {et~al.} 2025, \mnras, submitted (arXiv:2503.15590)

\bibitem[Dickey {et~al.}(2016)]{Dickey2016} Dickey, C.~M., van Dokkum, P.~G., Oesch, P.~A., {et~al.} 2016, \apj, 828, L11

\bibitem[Draine(2011)]{Draine2011} Draine, B.~T. 2011, Physics of the Interstellar and Intergalactic Medium (Princeton, NJ: Princeton Univ. Press)

\bibitem[Fakhouri {et~al.}(2010)]{Fakhouri2010} Fakhouri, O., Ma, C.~P., \& Boylan-Kolchin, M. 2010, \mnras, 406, 2267

\bibitem[Ferland {et~al.}(1998)]{Ferland1998} Ferland, G.~J., Korista, K.~T., Verner, D.~A., {et~al.}, 1998, PASP, 110, 761

\bibitem[Furtak {et~al.}(2024)]{Furtak2024} Furtak, L.~J., Labb\'e, I., Zitrin, A., {et~al.} 2024, Nature, 628, 57

\bibitem[Gaia Collaboration {et~al.}(2023)]{Gaia2023} Gaia Collaboration, Vallenari, A., Brown, A.~G.~A., {et~al.} 2023, \aap, 674, A1

\bibitem[Greene \& Ho(2005)]{Greene2005} Greene, J. E., \& Ho, L. C. 2005, ApJ, 630, 122

\bibitem[Greene {et~al.}(2024)]{Greene2024} Greene, J.~E., Labb\'e, I., Goulding, A.~D., {et~al.} 2024, \apj, 964, 39

\bibitem[Greene {et~al.}(2020)]{Greene2020} Greene, J. E., Strader, J., \& Ho, L. C. 2020, ARA\&A, 58, 257

\bibitem[Guia {et~al.}(2024)]{Guia2024} Guia, C.~A., Pacucci, F., Kocevski, D.~D. 2024, RNAAS, 8, 207

\bibitem[Hoffman \& Loeb (2007)]{Hoffman2007} Hoffman, L., \& Loeb, A. 2007, \mnras, 377, 957

\bibitem[Inayoshi(2025)]{Inayoshi2025a} Inayoshi, K. 2025, ApJ, submitted (arXiv:2503.05537)

\bibitem[Inayoshi et al.(2022)]{Inayoshi2022} Inayoshi, K., Harikane, Y., Inoue, A. K., Li, W., \&  Ho, L. C. 2022, \apj, 938, L10

\bibitem[Inayoshi \& Ichikawa(2024)]{Inayoshi2024} Inayoshi, K., \& Ichikawa, K. 2024, \apj, 973, L49

\bibitem[Inayoshi \& Maiolino(2025)]{Inayoshi2025b} Inayoshi, K., \& Maiolino, R. 2025, \apj, 980, L27

\bibitem[Ji {et~al.}(2025)]{Ji2025} Ji, X., Maiolino, R., \"{U}bler, H., {et~al.} 2025, \mnras, submitted (arXiv:2501.13082)

\bibitem[Jiang {et~al.}(2025)]{Jiang2025} Jiang, F., Jia, Z., Zheng, H., {et~al.} 2025, \apj, submitted (arXiv:2503.23710)

\bibitem[Kataria {et~al.}(2025)]{Kataria2025} Kataria, M., Saha, K., \& Elmegreen, B. 2025, \apj, in press (arXiv:2504.13664)

\bibitem[Killi {et~al.}(2024)]{Killi2024} Killi, M., Watson, D., Brammer, G., {et~al.} 2024, A\&A, 619, A52

\bibitem[Kocevski {et~al.}(2025)]{Kocevski2025} Kocevski, D.~D., Finkelstein, S.~L., Barro, G., {et~al.} 2025, \apj, in press (arXiv:2404.03576)

\bibitem[Kocevski {et~al.}(2023)]{Kocevski2023} Kocevski, D.~D., Onoue, M., Inayoshi, K., {et~al.} 2023, \apj, 954, L4

\bibitem[Kokorev {et~al.}(2024)]{Kokorev2024} Kokorev, V., Caputi, K.~I., Greene, J.~E., {et~al.} 2024, \apj, 968, 38

\bibitem[Koprowski {et~al.}(2024)]{Koprowski2024} Koprowski, M.~P., Wijesekera, J.~V., Dunlop, J.~S., {et~al.} 2024, A\&A, 691, A164

\bibitem[Kormendy \& Ho(2013)]{Kormendy2013} Kormendy, J., \& Ho, L. C. 2013, ARA\&A, 51, 511

\bibitem[Kroupa(2001)]{Kroupa2001} Kroupa, P. 2001, \mnras, 322, 231

\bibitem[Labb\'e {et~al.}(2025a)]{Labbe2025} Labb\'e, I., Greene, J.~E., Bezanson, {et~al.} 2025a, \apj, 978, 92

\bibitem[Labb\'e {et~al.}(2025b)]{Labbe2025b} Labb\'e, I., Greene, J.~E., Matthee, J., {et~al.} 2025b, \apj, submitted (arXiv:2412.04557)

\bibitem[Larson {et~al.}(2023)]{Larson2023} Larson, R., L., Hutchison, T., A., Bagley, M., {et~al.} 2023, \apj, 958, 141

\bibitem[Li {et~al.}(2025)]{Li2025} Li, Z., Inayoshi, K., Chen, K., {et~al.} 2025, \apj, 980, 36

\bibitem[Llerena {et~al.}(2024)]{Llerena2024} Llerena, M., Amor\'in, R., Pentericci, L., {et~al.} 2024, A\&A, 691, A59

\bibitem[Loveday {et~al.}(2012)]{Loveday2012} Loveday, J., Norberg, P., Baldry, I.~L., {et~al.} 2012, \mnras, 420, 1239

\bibitem[Ma {et~al.}(2025)]{Ma2025} Ma, Y., Greene, J.~E., Setton, D.~J., {et~al.} 2025, \apj, submitted (arXiv:2504.08032)

\bibitem[Madau \& Haardt(2024)]{Madau2024} Madau, P., \& Haardt, F. 2024, \apj, 976, L24

\bibitem[Maiolino {et~al.}(2024)]{Maiolino2024} Maiolino, R., Scholtz, J., Curtis-Lake, E., {et~al.} 2024, A\&A, 691, A145

\bibitem[Marchesini {et~al.}(2012)]{Marchesini2012} Marchesini, D., Stefanon, M., Brammer, G.~B., {et~al.} 2012, \aj, 748, 126

\bibitem[Matthee {et~al.}(2024)]{Matthee2024} Matthee, J., Naidu, R.~P., Brammer, G., {et~al.} 2024, \apj, 963, 129

\bibitem[Matthee {et~al.}(2025)]{Matthee2025} Matthee, J., Naidu, R.~P., Kotiwale, G., {et~al.} 2025, \apj, submitted (arXiv:2412.02846)

\bibitem[McQuinn(2016)]{McQuinn2016} McQuinn, M. 2016, ARA\&A, 54, 313

\bibitem[Meiksin(2006)]{Meiksin2006} Meiksin, A. 2006, \mnras, 365, 807

\bibitem[Morishita {et~al.}(2024)]{Morishita2024} Morishita, T., Stiavelli, M., Chary, R.-R., {et~al.} 2024, \apj, 963, 9

\bibitem[Nagao {et~al.}(2001)]{Nagao2001} Nagao, T., Murayama, T., Taniguchi, Y. 2001, \apj, 546, 744

\bibitem[Naidu {et~al.}(2025)]{Naidu2025} Naidu, R.~P., Matthee, J., Katz, H., {et~al.} 2025, Nature, submitted (arXiv:2503.16596)

\bibitem[Pacucci \& Narayan(2024)]{Pacucci2024} Pacucci, F., \& Narayan, R. 2024, \apj, 976, 96

\bibitem[P\'erez-Gonz\'alez {et~al.}(2024)]{Perez-Gonzalez2024} P\'erez-Gonz\'alez, P.~G., Barro, G., Rieke, G.~H., {et~al.} 2024, \apj, 968, 4

\bibitem[Perrin {et~al.}(2015)]{Perrin2015} Perrin, M., D., Long, J., Sivaramakrishnan, A., {et~al.} 2015, WebbPSF: James Webb Space Telescope PSF Simulation Tool, Astrophysics Source Code Library, record ascl:1504.007



\bibitem[Planck Collaboration {et~al.}(2020)]{Planck2018} Planck Collaboration, Aghanim, N., Akrami, Y., {et~al.} 2020, A\&A, 641, A6


\bibitem[Rinaldi {et~al.}(2025)]{Rinaldi2025} Rinaldi, P., Bonaventura, N., Rieke, G.~H., {et~al.} 2025, \apj, submitted (arXiv:2411.14383)

\bibitem[Rusakov {et~al.}(2025)]{Rusakov2025} Rusakov, V., Watson, D., Nikopoulos, G.~P., {et~al.} 2025, Nature, submitted (arXiv:2503.16595)

\bibitem[S\'ersic (1968)]{Sersic1968} S\'ersic, J. L. 1968, Atlas de Galaxias Australes (C\'ordoba: Obs. Astron., Univ. Nac. C\'ordoba)

\bibitem[Setton {et~al.}(2025)]{Setton2025} Setton, D.~J., Greene, J.~E., Spilker, J.~S., {et~al.} 2025, \apj, submitted (arXiv:2503.02059)

\bibitem[Simcoe {et~al.}(2004)]{Simcoe2004} Simcoe, R.~A., Sargent, W.~L.~W., \& Rauch, M. 2004, \apj, 606, 92

\bibitem[Solimano {et~al.}(2025)]{Solimano2025} Solimano, M., Gonz\'alez-L\'opez, J., \& Aravena, M., A\&A, 693, A70

\bibitem[Storey \& Zeippen (2020)]{Storey2000} Storey, P. J., \& Zeippen, C. J. 2000, \mnras, 312, 813

\bibitem[Suess {et~al.}(2024)]{Suess2024} Suess, K., Weaver, J.~R., Price, S.~H., {et~al.} 2024, ApJ, 976, 101

\bibitem[Tanaka {et~al.}(2025)]{Tanaka2025} Tanaka, T., Silverman, J.~D., Shimasaku, K., {et~al.} 2025, ApJ, submitted (arXiv:2412.14246)

\bibitem[Taylor {et~al.}(2025)]{Taylor2025} Taylor, A.~J., Finkelstein, S.~L., Kocevski, D.~D., {et~al.} 2025, ApJ, submitted (arXiv:2409.06772)

\bibitem[Tie {et~al.}(2024)]{Tie2024} Tie, S.~S., Hennawi, J.~F., Wang, F., {et~al.} 2024, MNRAS, 535, 223

\bibitem[van~der~Wel {et~al.}(2011)]{vanderWel2011} van der Wel, A., Straughn, A. N., Rix, H.-W., et al. 2011, \apj, 742, 111

\bibitem[Wang {et~al.}(1998)]{Wang1998} Wang, J., Heckman, T.~M., \& Lehnert, M.~D. 1998, \apj, 509, 93

\bibitem[Wang {et~al.}(2025)]{Wang2025} Wang, B., de Graaff, A., Davies, R.~L., {et~al.} 2025, \apj, in press (arXiv:2403.02304)

\bibitem[Wang {et~al.}(2024)]{Wang2024} Wang, B., Leja, J., De Graaff, A., {et~al.} 2024, \apj, 969, L13

\bibitem[Williams {et~al.}(2024)]{Williams2024} Williams, C.~C., Alberts, S., Ji, Z., {et~al.} 2024, \apj, 968, 34

\bibitem[Wu {et~al.}(2022)]{Wu2022} Wu, Z., Ho, L.~C., Zhuang, M.~Y. 2022, \apj, 941, 95

\bibitem[Zhang {et~al.}(2025)]{Zhang2025} Zhang, C., Wu, Q., Fan, X., et al. 2025, Nature Astronomy, submitted

\end{thebibliography}
\end{document}